\documentclass[iop]{emulateapj}

\usepackage{epstopdf} 

\newcommand{\kms}{\,km~s$^{-1}$}

\newcommand{\Msun}{\mbox{\,$M_{\odot}$}}
\newcommand{\Lsun}{\mbox{\,$L_{\odot}$}}

\def\spose#1{\hbox to 0pt{#1\hss}}
\def\simlt{\mathrel{\spose{\lower 3pt\hbox{$\mathchar"218$}}
     \raise 2.0pt\hbox{$\mathchar"13C$}}}
\def\simgt{\mathrel{\spose{\lower 3pt\hbox{$\mathchar"218$}}
     \raise 2.0pt\hbox{$\mathchar"13E$}}}

\shorttitle{M32 Stellar Kinematics}
\shortauthors{Howley~et~al.}

\begin{document}
 \title{
\vspace{0.05in}
Internal Stellar Kinematics of M32 from the SPLASH Survey:\\
Dark Halo Constraints and the Formation of Compact Elliptical Galaxies\altaffilmark{1}}

\author{
K.~M.\ Howley\altaffilmark{2}, 
P.\ Guhathakurta\altaffilmark{3}, 
R.\ van der Marel\altaffilmark{4},
M.\ Geha\altaffilmark{5}, 
J.\ Kalirai\altaffilmark{4}, \\
B.\ Yniguez\altaffilmark{6}, 
E.\ Kirby\altaffilmark{7,8}, 
J.-C.\ Cuillandre\altaffilmark{9},
and
K.\ Gilbert\altaffilmark{10,8}}

\altaffiltext{1}{Data herein were obtained at the W. M. Keck Observatory, which is operated as a scientific partnership among the California Institute of Technology, the University of California, and NASA. The Observatory was made possible by the generous financial support of the W. M. Keck Foundation.}
\altaffiltext{2}{Lawrence Livermore National Laboratory,
     P.O. Box 808, Livermore, CA~94551; {\tt howley1@llnl.gov}}   
\altaffiltext{3}{UCO/Lick Observatory, University of California
    Santa Cruz, 1156 High Street, Santa Cruz, CA~95064; {\tt raja@ucolick.org}}   
\altaffiltext{4}{Space Telescope Science Institute, 3700 San Martin Drive, Baltimore, MD~21218; {\tt [marel, jkalirai]@stsci.edu}}      
\altaffiltext{5}{Department of Astronomy, Yale University, New Haven, CT 06510; {\tt marla.geha@yale.edu}}  
\altaffiltext{6}{Physics \& Astronomy Department, University of California, Irvine, 4129 Frederick Reines Hall, Irvine, CA~92697  {\tt byniguez@uci.edu}}
\altaffiltext{7}{Department of Astronomy, California Institute of Technology, 1200 East California Blvd, Pasadena CA~91125 ; {\tt enk@astro.caltech.edu}} 
\altaffiltext{8}{Hubble Fellow}
\altaffiltext{9}{Canada-France-Hawaii Telescope, 65-1238 Mamalahoa Hwy, Kamuela, HI~96743; {\tt jcc@cfht.hawaii.edu}}
\altaffiltext{10}{Department of Astronomy, University of Washington, Box 351580, Seattle, WA~98195; {\tt kgilbert@astro.washington.edu}}  
 \begin{abstract}
As part of the SPLASH survey of the Andromeda galaxy (M31) and its neighbors,
we have obtained Keck/DEIMOS spectra of the compact elliptical (cE) satellite
M32.  This is the first resolved-star kinematical study of any cE galaxy.  In
contrast to previous studies that extended out to $r\lesssim30\arcsec\sim1\,r^{\rm eff}_I\sim100$~pc, we measure the rotation curve and velocity dispersion profile out to $r\sim250\arcsec$ and higher order Gauss-Hermite moments out to $r\sim70\arcsec$.   We achieve this
by combining integrated-light spectroscopy at small radii (where
crowding/blending are severe) with resolved stellar spectroscopy at larger
radii, using spatial and kinematical information to statistically account for
M31 contamination.
The rotation curve and
velocity dispersion profile extend well beyond the radius ($r\sim150\arcsec$)
where the isophotes are distorted.  Unlike NGC~205, another close dwarf
companion of M31, M32's kinematics are regular and symmetric and do not show
obvious sharp gradients across the region of isophotal elongation and twists.
We interpret M32's kinematics using three-integral axisymmetric dynamical
equilibrium models constructed using Schwarzschild's orbit superposition
technique.  Models with a constant mass-to-light ratio can fit the data
remarkably well.  However, since such a model requires an increasing
tangential anisotropy with radius, invoking the presence of an extended dark
halo may be more plausible.  Such an extended dark halo is definitely
required to bind a half-dozen fast-moving stars observed at the largest
radii, but these stars may not be an equilibrium component of M32.  The
observed regularity of the stellar kinematics, as well as the possible
detection of an extended dark halo, are unexpected if M31 tides are
significant at large radii.  While these findings by themselves do not rule
out tidal models for cE formation, they suggest that tidal stripping may not 
be as significant for shaping cE galaxies as has often been argued.
\end{abstract}
 
  \section{INTRODUCTION}\label{intro_sec}

Andromeda's (M31) nearest companion, M32, is our closest example of a compact elliptical (cE), a rare dwarf galaxy type.  Including M32, there are fewer than ten cE galaxies known within 100 Mpc \citep{roo65,kin73,smi09,chi09,chi10}. Compact elliptical galaxies have luminosities of $\sim10^9 \Lsun$, comparable to dwarf elliptical galaxies (dE), but with significantly smaller effective radii ($r_{\rm eff} \sim 0.1$--$0.7$ kpc), leading to notably higher surface brightnesses \citep[$\mu_{\rm eff} \sim 18$--$21$ mag arcsec$^{-2}$,][]{cho02,mie05,chi09}. While cEs do appear to be a rare galaxy type, the discovery of these objects has been slow due to their elliptical galaxy-like appearance at distances beyond the Local Group \citep{zie98,dri98}.

Compact elliptical galaxies are consistently found in close projection
to a massive parent galaxy 
\citep[$5 < r_{\rm proj} < 80$ kpc,][]{chi09}, indicating that gravitational effects play some role in
their evolution.  One formation scenario proposes that cEs are remnant
cores of tidally stripped `normal' galaxies \citep{fab73,nie90,bek01,gra02,chi10,hux10},
although the rarity of cEs relative to normal elliptical and spiral galaxies
suggests this does not happen often, or that cEs are short-lived.
Another formation theory suggests that cEs are low-mass classical
elliptical galaxies that were either captured by or formed in the
potential well of a massive neighbor 
\citep{nie87,bur94}.  This latter scenario is supported by the position of cEs on the
fundamental plane, at the low luminosity end of the classical
elliptical galaxy trend \citep{wir84,kor85,nie87}.
If these objects are indeed low mass normal
elliptical galaxies, then their rarity implies a steep fall off at the
faint end of the galaxy luminosity function 
\citep{bin88}.  
However, recent observational improvements indicate
that the light profiles of many cEs, including M32, are
better fit by a two component $r^{1/n}$ bulge + exponential profile,
characteristic of a disk galaxy, rather than the historically used
$r^{1/4}$ law, typical of elliptical galaxies \citep{gra02}.

M32 is the nearest example of the cE class and an excellent specimen for examining cE properties.
As is the case for other cEs, M32 lies
close to its parent galaxy at a projected separation of only 22\arcmin\ (5
kpc) from M31's center.  Our relative proximity to M32
allows us to resolve individual stars in this object --- at least for
outer radii where stellar crowding is less severe.  Photometric evidence
suggests that M32's physical distance from M31 is similar to its
projected distance: \citet{cho02} show that distortions in M32's
outer elliptical isophotes are consistent with the hypothesis that these
two galaxies are tidally interacting.  Tidal interactions are a possible
explanation for the galaxy's unusual stellar population gradient and
light profile \citep{fab73,oco80,dav00a,dav00b,bek01,gra02}.
Integrated light spectroscopic studies suggest the presence of a younger, more metal-rich stellar population at the center of M32
\citep[][and references therein]{ros05,coe09}
which is perhaps the result of a tidally induced nuclear starburst \citep{bek01}.  
 
Previous studies of M32's internal kinematics have primarily focused
on the inner regions.  The steep rotational velocity gradient at
the center of M32 indicates the presence of a central dark mass.
High resolution imaging and integrated light spectroscopy has confirmed the presence of a central black
hole with mass $M_{\rm BH} \sim 2$--$4 \times 10^6$ \Msun 
\citep{goo89,ben96,van98b,jos01,ver02,tre02,van97,van10}. While these studies have been appropriate for determining the mass of
the central dark object, their limited radial extent (generally $r < 10\arcsec$) does not
provide much information on the wider galaxy environment.  Past
attempts to study the kinematics of M32 at larger radii ($r < 60\arcsec$) have
produced conflicting results 
 \citep{ton84,car93}.  Specifically, \citet{ton84} found the velocity dispersion to increase
outwards while \citet{car93} found it to be decrease outwards.  Both authors profess
problems in their measurements due to complicated sky subtraction and, in
particular, contamination from M31.

In this paper, we use a combination of integrated light and resolved
stellar spectroscopy to obtain an accurate kinematical profile of
M32 out to $\sim8 r_{\rm eff}$ --- a much larger radius than has been previously
possible.  In the inner regions of M32, crowding is so significant
that individual stars cannot be resolved, while in the outer regions,
the integrated light spectrum becomes very noisy due to the presence
of M31 and the steep light profile of M32.  This is the first 
attempt to combine spectra of individual stars with integrated
light spectroscopy to obtain a complete picture of M32.  
This work is part of the SPLASH Survey (Spectroscopic and Photometric Landscape of Andromeda's Stellar Halo), aimed at the study of M31 and its satellites.

This paper is organized as follows.  
In \S\,\ref{sec_spec}, we measure the internal kinematics of M32 out to a radius of $\sim8r_{\rm{eff}}$ ($\sim 1$~kpc) using resolved stellar light and $\sim3r_{\rm{eff}}$ ($\sim0.3$~kpc) using integrated light.  We then present velocity, velocity dispersion and Gauss-Hermite moment major- and minor- axis profiles from these measurements and compare to previous measurements.
Using these results, we construct an axisymmetric,
three-integral model in \S\,\ref{sec_analysis} to obtain an estimate of M32's mass and
M/L. Finally, in \S\,\ref{sec_discussion} we summarize and discuss our results.

Throughout this paper, we assume that M32 has the same distance modulus as
M31: $(m-M)_0=20.54\pm0.03$ \citep[$785\pm25$~kpc,][]{mcc05}.  At M32/M31's
distance, 1\arcsec\ is equivalent to 3.8 pc.  While the distance to M32 has
been measured independently \citep{jen03,kar04}, it is consistent with the
distance to M31.  Moreover, the M32 distance estimate is affected by crowding
problems and M31 contamination.  The lack of obvious signs of dust extinction
in M32 suggests that the satellite lies in front of M31's disk \citep{for78},
but a precise M31-M32 distance is yet to be established.

\section{STELLAR KINEMATICS OF M32}\label{sec_spec}

In this section, we give a detailed account of our M32 kinematical measurements.
In \S\,\ref{ssec_resolve} we present measurements from the resolved stellar kinematics. In \S\,\ref{ssec_integrated} we present measurements from integrated light.  For readers not interested in these details, 
an integrated view of M32's kinematics and a comparison to previous studies are presented in \S\,\ref{ssec_vprofile}.

\subsection{Resolved Stellar Population Spectroscopy}\label{ssec_resolve}

In this subsection, we discuss our observational setup, including photometric and astrometric measurements (\S\,\ref{sssec_imaging}), mask placement (\S\,\ref{sssec_mask}), identification of isolated sources (\S\,\ref{sssec_target}), target selection and slit mask design (\S\,\ref{sssec_design}), and observing details (\S\,\ref{sssec_multi}).
We then discuss the reduction and analysis of the spectra, including the data reduction process (\S\,\ref{sssec_mreduce}), measurement of individual stellar velocities (\S\,\ref{sssec_cc}), quality assessment (\S\,\ref{sssec_quality}), and detection of and velocity measurements for serendipitously detected stars (\S\,\ref{sssec_serendip}).
Finallly, in  \S\,\ref{sssec_ml} we make velocity and velocity dispersion measurements along M32's major and minor axes via maximum likelihood analysis.

  \subsubsection{Pre-Imaging}\label{sssec_imaging}

We derive photometric and astrometric catalogs of stars in M32's general
vicinity from archival data obtained with the MegaCam\footnote{{\tt
http://www.cfht.hawaii.edu/Instruments/Imaging/MegaPrime/};
MegaPrime/MegaCam is a joint project of CFHT and CEA/DAPNIA at the
Canada-France-Hawaii Telescope (CFHT) which is operated by the National
Research Council (NRC) of Canada, the Institut National des Science de
l'Univers of the Centre National de la Recherche Scientifique (CNRS) of
France, and the University of Hawaii.  The observations were obtained by the
MegaCam instrument team in November 2004.}
imager on the 3.6~m Canada-France-Hawaii Telescope (CFHT).  These data are in
the form of $2^\circ\times2^\circ$ mosaic images in the $g^\prime$,
$r^\prime$, and $i^\prime$ bands centered on M31.

The mosaic image in each band was built from four separate MegaCam pointings each with a field of view of $1^\circ\times1^\circ$.
Each MegaCam pointing consists of five dithered, slightly overlapping
exposures with integration times of $5\times45$\,s in each of the $g^\prime$ and $r^\prime$ bands
and $5\times90$\,s  in the $i^\prime$ band.  Thus, the effective integration times are
225\,s in each of the $g^\prime$ and $r^\prime$ bands and 450\,s in the $i^\prime$ band.
$r^\prime$ bands, and 450\,s in the $i^\prime$ band.  

The data were
obtained under photometric conditions.  Individual exposures were detrended
using CFHT's data reduction pipeline, Elixir, and stacked into mosaic images
using the SWarp software \citep{ber02}.  The pixel scale on the mosaic image
is 0\farcs36, a factor of 2 coarser than the native MegaCam pixel scale.  The
median seeing FWHM is 1\farcs2, 1\farcs0 and 0\farcs7\ in $g^{\prime}$,
$r^{\prime}$, and $i^{\prime}$, respectively.  In order to highlight
point-like sources, a high-pass filtered image is constructed by subtracting
a smoothed version of the mosaic image from the original.

\begin{figure*}
\plotone{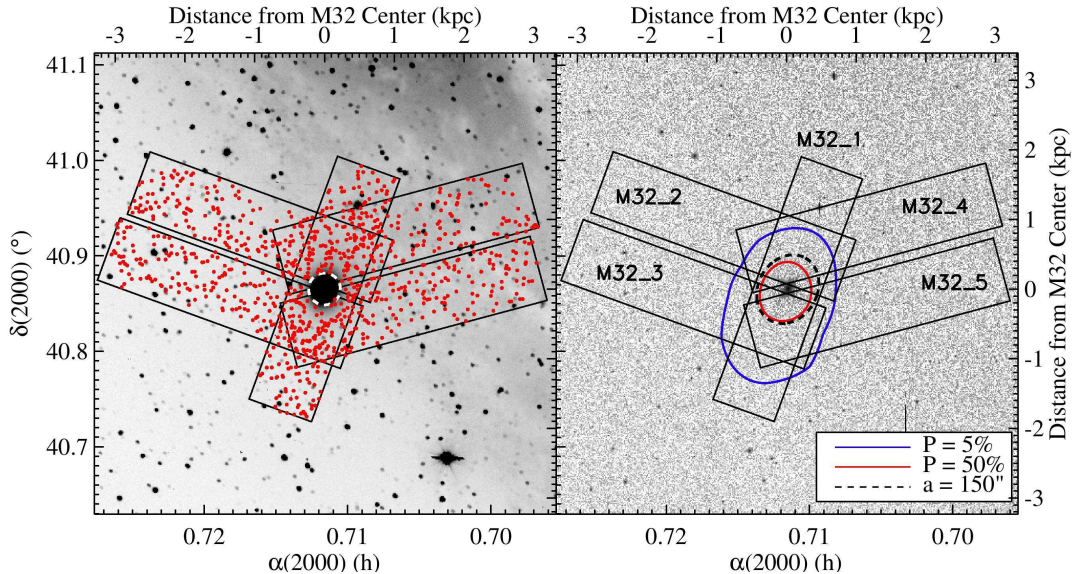}
\caption[Positions of the five observed Keck/DEIMOS multislit masks]{
Positions of the five observed Keck/DEIMOS multislit masks (M32\_1 -- M32\_5, black rectangles) shown within a $36^{\prime} \times 36^{\prime}$ section centered on M32.    Each DEIMOS mask is $\sim16\arcmin \times 4\arcmin$ 
\textbf{Left:} Masks overlaid on the \citet{cho02} KPNO Burrell Schmidt $B$-band image.  Red circles mark the location of our spectroscopic targets.  The dashed white circular contour at the center of M32 denotes the $i^\prime$ surface brightness magnitude limit at which crowding becomes so significant that individual targets can no longer be resolved in the CFHT/MegaCam image ($\mu_{i^{\prime}} \sim 19\ {\rm mag/arcsec}^2$, $r_{\rm M32}\sim 1\arcmin$). The surface brightness gradient seen in this image (brighter towards the north-west) is contaminating light from M31.  
\textbf{Right:} Masks overlaid on the high-pass filtered CFHT/MegaCam image
used for target selection.  The dashed black line marks the location where
M32's isophotes begin to show distortion \citep{cho02}.  The elliptical-like
blue and red contours mark the location of M32's 5\% and 50\% membership
probabilities, respectively, determined on the basis of model fits to the 2D
surface brightness distributions of M31's bulge/disk and M32.  
}
\label{fig_masks}
\end{figure*}

For the purposes of this M32 study, we analyze a $36^\prime\times36^\prime$
section of the $i^\prime$ and $r^{\prime}$ mosaic images centered on M32
(given the high degree of crowding in this region, the $g^{\prime}$ image
is not particularly useful due to its relatively poor seeing and the $r^{\prime}$ image is only used in the construction of Figure \ref{fig_cmd}; neither the $g^{\prime}$ nor $r^{\prime}$ images are used for spectroscopic selection).
The size of this image section is chosen to comfortably allow for the placement of
multiple Keck/DEIMOS multislit spectroscopic masks (each mask covers about
$16^\prime\times4^\prime$), as shown in Figure~\ref{fig_masks}.  

\begin{figure*}\plotone{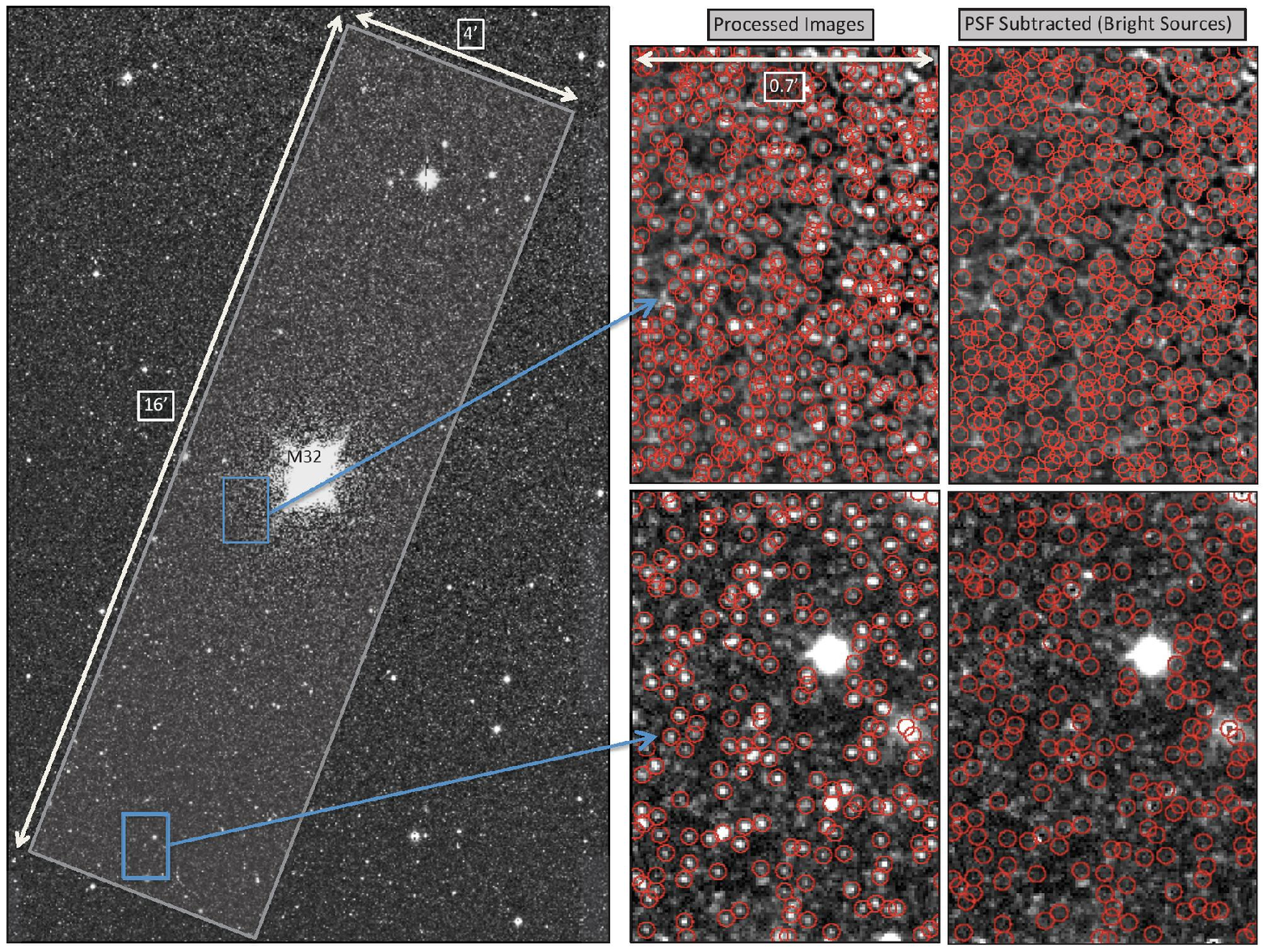}
\caption[Example of the CFHT/MegaCam image quality in various regions.]{Example of the CFHT/MegaCam image quality in various regions.  \textbf{Left:} Outline of the M32\_1 DEIMOS mask (grey) on the high-pass filtered CFHT/MegaCam image with three $0\farcs7 \times 1\farcs0$ blue boxes defining representative areas zoomed into on the right.  \textbf{Middle:} The column of plots demonstrates the quality of the image and the density of targets at the various locations on the mask.  The selections show: (top) region near M32, and (bottom) the southern part of the mask.  The red circles mark the location of the sources detected in the photometry.  \textbf{Right:}  The PSF subtracted images. 
}
\label{fig_cfht}
\end{figure*}

\begin{figure*}\plotone{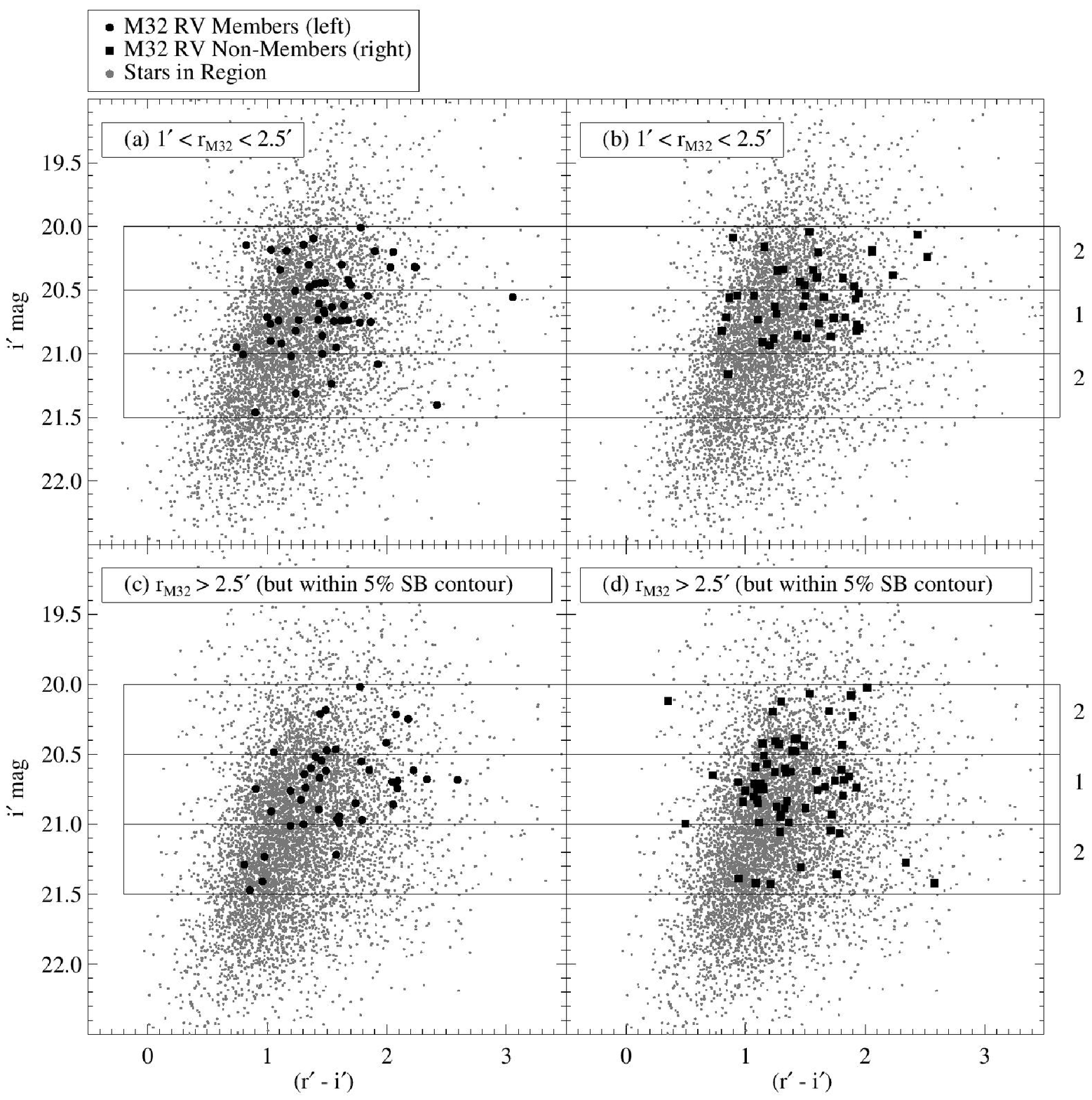}
\caption[Color-magnitude diagram (CMD) in the M32 region from a high-pass filtered CFHT MegaCam image (for illustration only).]{Color-magnitude diagram (CMD) based on the $i^\prime$ and $r^\prime$ photometry from the high-pass filtered CFHT MegaCam image (for illustration only due to the poor quality of the $r^\prime$ data).  Stars shown are within the $5\%$ M32 membership probability contour (see Figure~\ref{fig_masks}, right) and outside $1\arcmin$ of M32's severely crowded center (see Figure~\ref{fig_masks}, left).  The absolute calibration of the data is approximate and based on the TRGB magnitudes for M31 from \citet{mcc05}.  The black circles/squares denote Keck/DEIMOS spectroscopic targets with measured velocities. The numbered rectangular outlines mark the location of our spectroscopic priorities: (1) primary and (2) secondary. 
\textbf{Top~Left:} Black circles show M32 candidate members (i.e. stars with velocities in the range $-275 \leq v \leq -125$\kms) within $1\arcmin < r_{\rm M32} < 2\farcm5$. 
\textbf{Top~Right:} Black squares show M32 non-members (i.e. stars with velocities in the range $v < -275$\kms\ or  $v> -125$\kms) within $1\arcmin < r_{\rm M32} < 2\farcm5$.  
\textbf{Bottom~Left:} Black circles show M32 candidate members with $r_{\rm M32} > 2\farcm5$ (but within M32's $5\%$ probability contour, i.e. $r_{\rm M32}< 5\farcm3$). 
\textbf{Bottom~Right:} Black squares show M32 non-members with $r_{\rm M32} > 2\farcm5$ (but within M32's $5\%$ probability contour, i.e. $r_{\rm M32}< 5\farcm3$).  
Note the similarity between all four populations, thereby illustrating the difficulty of photometrically disentangling these two populations.}
\label{fig_cmd}
\end{figure*}

We generate star lists over the high-pass filtered $r^\prime$ and $i^\prime$
MegaCam mosaic images using the FIND task in the DAOPHOT photometry package
\citep{ste94} and carry out preliminary aperture photometry of all detected
sources.  A spatially varying point spread function (PSF) template is
iteratively derived from a set of bright, isolated stars whose neighbors have
been subtracted using these star lists.  This PSF template is then fit to all
sources in the catalog for each frame using the ALLSTAR module to produce
accurate photometric catalogs.  Best fit PSF templates are then subtracted
from the high-pass filtered image to create a subtracted image that shows the
residuals due to imperfect PSF subtraction and missed sources.  The source
find procedure and photometry procedures are then repeated on the PSF
subtracted images several times to identify faint objects missed in earlier
passes, and this yields a final photometric catalog of over $10^5$ stars.
Next, the magnitudes are roughly calibrated based on the tip of the red giant
branch (TRGB) in each filter; precise photometric calibration is not needed
for this project since the purpose of our photometric catalog is solely to
assign rough $i^\prime$ magnitude ranges from which to select and set
priorities for spectroscopic targets.  As a final step, we astrometrically
correct all of our $x$ and $y$ positions by computing transformations based
on the USNO A2.0 catalog.  The final RA and DEC of all sources are accurate
to $\sim$0.2\arcsec.  We demonstrate our photometric characterization of
sources in the $i^\prime$ image by illustrating both original and PSF
subtracted images in the vicinity of M32 in Figure~\ref{fig_cfht}.  

Figure~\ref{fig_cmd} shows the $i^\prime$ and $r^\prime$ photometry for stars within the $5\%$ membership probability contour centered on M32 (see Figure~\ref{fig_masks}, right).  Note that the color-magnitude diagrams (CMDs) in Figure~\ref{fig_cmd} are limited by the poor quality of the $r^\prime$ data, and are therefore used here for illustration purposes only.  The plots on the left show the photometry of targets in the velocity range of M32 ($-275 < v < -125$\kms, see \S\,\ref{sssec_ml}), while the figures on the right plot the remaining non-M32 like targets. All four CMDs look very similar,  thus illustrating the difficulty of photometrically preselecting stars that are likely to be M32 members.  At best, the CMDs can be used to preselect stars at or below the TRGB ($i^\prime \geq 20$), and above our spectroscopic limit ($i^\prime \leq 22$).

\subsubsection{Mask Placement}\label{sssec_mask}

The two largest obstacles to spectroscopy of the resolved stellar population of M32: contamination by M31 and crowding.  We consider each of these in turn.

M32 is projected against the high surface brightness inner regions of M31.  The probability of a star being an M32 member can be estimated by the M32/M31 surface brightness ratio at that location.  The two-dimensional surface brightness of M32 is modeled as a series of concentric ellipses based on the \citet{cho02} measurements of $I$-band surface brightness, ellipticity and position angle as a function of radius.  The two-dimensional surface brightness of M31 is modeled as the sum of a Sers\'ic bulge and an exponential disk based on published $V$-band surface brightness data and a mean color $V-I=1.6$ \citep{wal87,pri94,guh05,irw05}.  The $5\%$ and $50\%$ M32/(M31$+$M32) $I$-band surface brightness ratios are shown in Figure~\ref{fig_masks} (for a more detailed mapping see Figure~\ref{fig_stars}).  The probability of M32 membership drops off very rapidly with increasing radius because of M32's steep surface brightness profile.

Near the center of M32, however, it is difficult, if not impossible, to find isolated spectroscopic targets in the high-pass filtered CFHT image.  Even the most luminous RGB stars (that are otherwise ideal for spectroscopy) are badly blended.  Based on visual inspection (see \S\,\ref{sssec_target}), we avoid the inner $r_{\rm M32} \lesssim 1\arcmin$ region, corresponding to a surface brightness of $\mu_{I}=18.9$ mag arcsec$^{-2}$.

The arrangement of our five Keck/DEIMOS multislit masks  is shown in Figure~\ref{fig_masks}.  Each mask covers $\approx 16\arcmin \times 4\arcmin$.   The masks provide coverage of the $r_{\rm M32}<150\arcsec$ region where M32's isophotes are regular, and the $r_{\rm M32}>150\arcsec$ region where isophotal distortion is observed \citep{cho02}.  The first slitmask, M32\_1, is centered on M32 with the long axis rotated to a position angle $\phi$ of $160^\circ$, i.e. the approximate position angle of the inner elliptical isophotes \citep{cho02}.   The remaining four multislit masks, M32\_2$-$M32\_5, are oriented to optimize coverage of the outer regions.  M32's steep brightness profile implies that there is only a limited region in which neither contamination nor crowding is too severe; our arrangement of masks ensures good coverage of this region, while also covering M31's inner spheroid and disk for other SPLASH survey science. 

\subsubsection{Identifying Isolated Sources}\label{sssec_target}

Our chosen placement of the five Keck/DEIMOS masks defines the footprint over which spectroscopic targets are selected.  We next identify stars within this footprint which are least affected by crowding/blending. We use two criteria to reject targets: (1) cases where DAOPHOT finds one or more bright neighbors that are close to, but distinct from, the target (we refer to this as \textit{crowding}), and (2) cases where an apparent single source in the DAOPHOT catalog is a poor fit to the PSF (we refer to this as \textit{blending}). 

We address the issue of crowding by rejecting a target star if it has a neighbor in the DAOPHOT catalog that is so close/bright that the PSF of the neighbor significantly overlaps that of the target.  Based on visual inspection of crowded regions of the image, we have come up with an empirical criterion.  Any target that has even a single neighbor satisfying the following relation is eliminated from the list of potential spectroscopic targets:
\begin{equation}
{I}_{\rm tgt} >  {I}_{\rm nbr} + \frac{|\vec{r}_{\rm tgt,nbr}|}{0.8\arcsec} - 3.0
\end{equation}
where $I_{\rm tgt}$ and $I_{\rm nbr}$ are the apparent magnitudes of the target and neighbor, respectively, and $\vec{r}_{\rm tgt,nbr}$ is the projected distance between the two objects. Of the sources in the DAOPHOT catalog in the magnitude range $I=20-22$ (the range used to select spectroscopic targets, see \S\,\ref{sssec_design}), $\approx10$\% pass this crowding test; the surviving fraction increases with target brightness over this magnitude range.

We address the issue of blending by visually inspecting the images at the locations of the stars that survive the crowding test.  This inspection includes both the high-pass filtered and PSF-subtracted versions of the $i^\prime$-band CFHT/MegaCam image, as illustrated in Figure~\ref{fig_cfht}.  Each target is flagged as unblended (high priority), marginally blended (medium priority), or badly blended (low priority), depending on the degree to which its image resembles the PSF on the high-pass filtered image and the strength of systematic residuals at its location on the PSF-subtracted image. 

These de-blending exercises are only good to a point as we are limited by the $0\farcs8$ seeing (FWHM) of the CFHT/MegaCam image. The seeing FWHM was significantly better than this during the Keck/DEIMOS spectroscopic observations (\S\,\ref{sssec_multi}).  As a result, further de-blending of sources is carried out in the spatial and spectral domains, as discussed in \S\,\ref{sssec_serendip}.  See \citet{dor12} for a discussion of an automated procedure for
identifying blended sources in this data set.

\subsubsection{Mask Design}\label{sssec_design}

Targets are prioritized for spectroscopic observation based on two criteria: level of blending and magnitude.  The first prioritization, based on level of blending, divides the targets into three lists: list 1 -- unblended, list 2 -- marginally blended, and list 3 -- badly blended (as discussed in \S\,\ref{sssec_target}).  The second prioritization, based on magnitude, assigns priorities within each list based on M31's TRGB magnitude of $20.5$ \citep{mcc05}; the highest priority targets (priority 1) have  $I=20.5-21$, and the lowest priority targets (priority 2) have $I=20-20.5$ or $I=21-21.5$.  Targets with magnitudes outside the range $I=20-21.5$ are excluded.  The distribution of targets across these magnitude ranges is shown in Figure~\ref{fig_cmd}.

Five Keck/DEIMOS multislit masks are designed using A.~C. Phillip's {\tt dsimulator} software\footnote{\tt http://www.ucolick.org/$\sim$phillips/deimos\_ref/masks.html}.  The software takes as input the multiple target lists (lists 1--3) organized by target priority (priorities 1--2).  Each $\approx 16\arcmin \times 4\arcmin$ Keck/DEIMOS mask is populated with targets from list 1, in order of decreasing priority, followed by list 2, and so on.  Our selection process is identical to that discussed in Appendix A of \citet{guh06}, with the following modifications: target prioritization, minimum distance between target and slit end ($1\farcs65$), and distance between adjacent slitlets ($0\farcs3$).  The location of the slitlets selected for observation is shown in Figure~\ref{fig_masks}.  The five masks contain a total of 883 slitlets.
\subsubsection{Observations}\label{sssec_multi}
 
Five multislit masks in the region of M32 were observed between November 2007 and August 2008 using the DEIMOS spectrograph \citep{fab03} on the Keck II 10~m telescope.  The arrangement of the masks is shown in Figure~\ref{fig_masks}.  The observing details are summarized in Table \ref{tab_obs}.

\begin{deluxetable*}{lllccccccc}
\tabletypesize{\scriptsize}
\tablecaption{Keck/DEIMOS Multislit Mask Exposures}
\tablewidth{0 pt}
\tablehead{
\colhead{Mask}&
\colhead{Observation}&
\colhead{$\alpha$}&
\colhead{$\delta$}&
\colhead{P.A.}&
\colhead{t$_{\rm{exp}}$}&
\colhead{Seeing}&
\colhead{No. of}&
\colhead{No. of Useable}&
\colhead{No. of Useable}\\
\colhead{Name}&
\colhead{Date}&
\colhead{(J2000.0)}&
\colhead{(J2000.0)}&
\colhead{(deg)}&
\colhead{(m)}&
\colhead{FWHM}&
\colhead{Slits}&
\colhead{Target Velocities}&
\colhead{Serendip Velocities}
}
\startdata
	M32\_1\tablenotemark{a} &  2007 Nov 14 & 00 42 38.3 & +40 51 34.0  & 160 
	& $2 \times 20$ & $0\farcs5$ & 194 & 189 (97\%) & 73\\
	M32\_2 &  2008 Aug 03 & 00 43 03.8 & +40 55 07.7 & 70
	& $3 \times 20$ & $0\farcs6$ & 184 & 169 (92\%) & 17\\
	M32\_3 &  2008 Aug 03 & 00 43 11.6 & +40 52 34.7 & $-$110 
	& $3 \times 20$ & $0\farcs7$  & 191 & \hspace{0.025in} 134 (70\%)\tablenotemark{b} & 10\\
	M32\_4 & 2008 Aug 04 & 00 42 13.9 & +40 54 44.2 & 105 
	& $3 \times 20$  & $0\farcs6$ & 143 & 137 (96\%) & 117\\
	M32\_5 &  2008 Aug 04 & 00 42 13.9 & +40 52 02.6 & $-$75 
	& $3 \times 20$  & $0\farcs6$ & 171 & 157 (92\%) & 81 \\[1 mm]
\tableline\tableline \vspace{1 mm}
	Total: &&&&&&& 883 & 786 (89\%)& 298 
\enddata
\tablecomments{Units of right ascension ($\alpha$) are in hours, minutes and seconds.  Units of declination ($\delta$) are in degrees, arcminutes and seconds.}
\tablenotetext{a}{The ``M32\_1" mask was originally named ``M32" at the time of submission of the mask design.}
\tablenotetext{b}{Buckling of the M32\_3 DEIMOS multislit mask at time of observation adversely
affected $\approx25\%$ of its slitlets.}
\label{tab_obs}
\end{deluxetable*}

All five multislit masks were observed with the 1200 line mm$^{-1}$ grating.  This configuration yields a spatial scale of $0\farcs12$ pixel$^{-1}$ and a spectral dispersion of $0.33$~\AA~pixel$^{-1}$.  We set the central wavelength to $7800$ \AA,  corresponding to wavelength range of $\sim 6450-9150$ \AA.  The exact wavelength range for each slit varies as a result of location on the multislit mask and/or truncation due to vignetting.  
The wavelength region is chosen to target several spectral features including the strong 
Ca {\scriptsize\rm{II}} triplet absorption feature present in RGB stars.  The anamorphic distortion factor for this grating and central wavelength is 0.606. Therefore, the $0\farcs8$ wide slitlets subtend $4.1$ pixels.  Better still, excellent seeing conditions ($\sim0\farcs6$) during observations provide somewhat better spectral resolution yielding an average resolution of 3.1~pixels~=~1.0~\AA\ FWHM. 

Useful spectra are obtained from 786 of the 883 slitlets ($89\%$).
The success rate would have been even higher were it not for the fact that $\approx25\%$
of the slitlets on mask M32\_3 were lost due to buckling of the mask during its insertion
into the DEIMOS focal plane at the time of observations; fortunately the buckling took
place at the ENE end of the mask away from M32.

\subsubsection{Data Reduction}\label{sssec_mreduce}
 
The five Keck/DEIMOS multislit masks are processed using the \textit{spec2d} and \textit{spec1d} software (version 1.1.4) developed by the DEEP Galaxy Redshift Survey team at the University of California, Berkeley\footnote{\tt http://astro.berkeley.edu/$^{\sim}$cooper/deep/spec2d/}.  Briefly, the reduction pipeline rectifies, flat-field and fringe corrects, wavelength calibrates, sky subtracts, and cosmic ray cleans the two-dimensional spectra, and extracts the one-dimensional spectra.  We give a more detailed description of the reduction process below.  
 
 First, the reduction pipeline rectifies curved spectra into rectangular arrays by applying small shifts and interpolating in the spatial direction.  One-dimensional slit function, two-dimensional flat-field and fringing corrections are then applied to the spectra.  The wavelength solution of the rectified spectra is obtained by fitting a polynomial to the arc lamp lines (precise at the 0.01\AA\ level).   The two-dimensional spectra are then sky subtracted and cosmic ray cleaned. Sky is identified by collapsing the two-dimensional spectra in the wavelength direction to locate spatial positions along the slit that are least affected/unaffected by targets and serendipitous sources (see \S\,\ref{sssec_serendip} for a discussion of serendipitous sources).  Each two-dimensional spectrum is sky subtracted by fitting a B-spline model (wavelength as a function of two-dimensional position: $x$, $y$) to the night sky emission lines in the baseline portion of the spatial intensity profile.  This careful sky subtraction is of particular importance around the Ca {\scriptsize\rm{II}} triplet region due to the presence of bright night sky lines; poor sky subtraction would reduce our ability to accurately measure stellar velocities.  The two-dimensional exposures are then combined along with cosmic ray rejection and inverse variance weighting to create a single mean two-dimensional spectrum for each slit.  

\begin{figure*}
\plotone{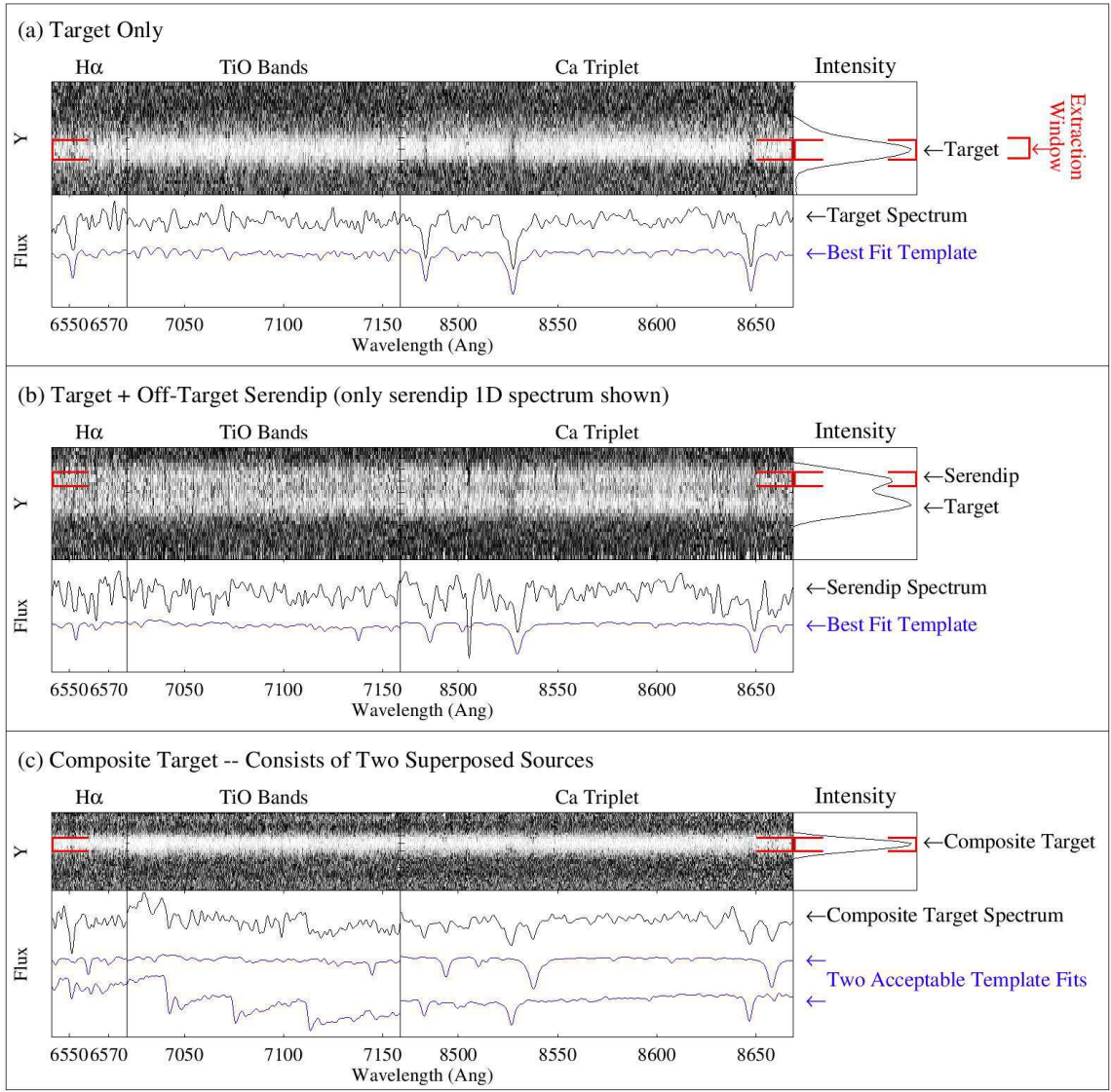}
\caption[Examples of target stars and serendipitous sources (\textit{serendips}).]{
Examples of target stars and serendipitous sources (\textit{serendips}).  Each panel (a--c) shows a
slitlet's two-dimensional spectrum (top left) where the wavelength axis $x$ runs horizontally (zoomed in to highlight specific spectral features), and the spatial axis $y$ (position along the slitlet) runs vertically.  The horizontal streak(s) within each two-dimensional spectrum is the stellar continuum from RGB candidate(s).  The spatial location of continuum is determined by collapsing the two-dimensional spectrum in the wavelength direction and identifying peaks in the intensity profile (top right, black).  One-dimensional stellar spectra (bottom, black) are extracted from the two-dimensional spectrum using a small spatial extraction window (top, red) that straddles the continuum of the desired star.  The LOS velocity of a star is measured by cross-correlating Doppler shifted stellar templates with the one-dimensional stellar spectrum until a best match is found (bottom, blue).  Note that the best-fit Doppler shifted stellar template (blue) is offset from the stellar spectrum (black) for the purpose of illustration. 
\textbf{(a)} Example of a slitlet intersecting a single star (i.e. the target star).
\textbf{(b)} Example of a slitlet intersecting two spatially resolved stars (i.e. the target star and an \textit{off-target serendip}).   The one-dimensional spectrum and best-fit stellar template shown correspond to the \textit{serendip}.
\textbf{(c)} Example of a slitlet intersecting two spatially blended stars (i.e. the target star and an \textit{on-target serendip}).  The presence of multiple stars is revealed by the two sets of absorption features in the one-dimensional spectrum.   
}
\label{fig_spec}
\end{figure*}

Last, the target is identified and its one-dimensional spectrum extracted.  Targets are located on the two-dimensional spectrum by identifying the peak brightness distribution in spatial intensity profile, obtained by collapsing the two-dimensional spectrum in the wavelength direction.  Target one-dimensional spectra are extracted from the two-dimensional spectra using a small spatial extraction window centered on the target. The one-dimensional spectra are re-binned into logarithmic wavelength bins with 13.8\kms/pixel.  The final result is a wavelength calibrated, sky subtracted, cosmic ray cleaned one-dimensional spectrum for each target.  An illustration of this process is shown in Figure~\ref{fig_spec}a.

 \subsubsection{Cross Correlation Analysis}\label{sssec_cc}

Line-of-sight (LOS) velocities for resolved targets are measured from the one-dimensional spectra using a \citet{sim07} modified version of the visual inspection software \textit{zspec}, developed by D. Madgwick  for the DEEP Galaxy Redshift Survey at the University of California, Berkeley.  The software determines the best-fit LOS velocity for a target by cross correlating its one-dimensional science spectrum with high S/N stellar templates in pixel space and locating the best-fit in reduced $\chi^2$ space.  The ten best-fit templates, LOS velocities, reduced-$\chi^2$ values and cross-correlation errors are reported. The stellar templates used in the cross-correlation analysis cover a wide range of stellar types, F8 to M8 giants, subgiants and dwarf stars, and metallicities, [Fe/H] $= -2.12$ to +0.11 \citep{sim07}.  The observing setup for the templates is nearly identical to that discussed in \S\ \ref{sssec_multi}, with the exception that template stars are observed with $0\farcs7$ wide slitlets, the minimum slit length is set to $4\arcsec$ (to allow for adequate sky subtraction), and the template stars are trailed across the slit. 

A-band telluric corrections and heliocentric corrections are calculated and applied to the measured LOS velocities.  The A-band telluric corrections, which account for velocity errors associated with the slight mis-centering of a star in a slit, are determined using the method discussed in \citet{soh07}. 

LOS velocity errors are determined for each star by scaling the
cross-correlation based velocity error using 
duplicate radial velocity measurements of stars \citet[][K.~Gilbert, private communication].  The average LOS velocity error for M32 RGB
The velocity error for
each star $\Delta v$ is estimated to be:
\begin{equation}
\label{eqn_dv}
\Delta v = \sqrt{(1.85\times \Delta v_{\rm{cc}})^{2} + 2.2^{2}}
\end{equation}
where $\Delta v_{\rm cc}$ is the cross-correlation based error and 2.2\kms\,
is the systematic velocity error as determined by \citet{sim07}.  The scale
factor 1.85 is determined from duplicate radial velocity measurements of
stars. The average LOS velocity error for M32 RGB stars is 4\kms.

\subsubsection{Quality Assessment}\label{sssec_quality}

Each two-dimensional spectrum, one-dimensional spectrum, and corresponding Doppler shifted template match, are visually inspected in \textit{zspec} and assigned a quality code $Q$ based on the reliability of the fit.  This process allows the user to judge the quality of a spectrum and reject instrumental failures and poor quality spectra.   Velocity measurements based on two or more strong spectral features are assigned $Q=4$ ($82\%$ of targets).  Velocity measurements based on one strong feature plus additional marginal features are assigned $Q=3$ ($6\%$ of targets).  Spectra that contain no strong features, low S/N and/or instrumental failures are assigned $Q=2$ ($8\%$ of targets).  For cases in which \textit{zspec} did not return an accurate velocity measurement, but visual inspection of the one-dimensional spectrum showed an obvious velocity shift, the velocity is manually marked and assigned $Q=1$ ($<1\%$ of targets).  Foreground stars  used for the purpose of alignment are assigned $Q=-1$ ($3\%$ of targets).  Additional details on this quality code assignment can be found in \citet{guh06}.

\subsubsection{Serendipitous Sources}\label{sssec_serendip}

Upon visual inspection of the one-dimensional and two-dimensional spectra during the quality assessment phase outlined in \S\,\ref{sssec_quality}, some fraction of the slits clearly show that the full length of the slitlet intersects more than one star: the target star and one or more serendipitously detected stars, known hereafter as a \textit{serendips}. These detections occur frequently in our target region due to the severe crowding and blending in the CFHT MegaCam data.   It is common for the full length of the slit to intersect multiple isolated sources in addition to the target (i.e. neighbors) as a result of the severe crowding in the region.  Sources that are nearly blended with the target are also commonly found; this is in part because the de-blending exercises for target selection (see \S\,\ref{sssec_target}) are good only to the $0\farcs8$ seeing limit of the CFHT MegaCam data.  The better angular resolution of the Keck/DEIMOS spectroscopic data ($\sim 0\farcs6$) allows for the spatial resolution of stars that can not be resolved in the CFHT MegaCam data. 

\textit{Serendips} are detected via one of two methods: through continuum detections that are offset from the primary target in the spatial direction (referred to as \textit{off-target serendips}), or by the detection of spectral features that are offset from the primary target in the spectral direction (referred to as \textit{on-target serendips}).  In a couple rare cases, we detect stars using a combination of the two methods (referred to as \textit{off-target superimposed serendips}).  We discuss the details of these detection methods in turn.  

\textit{Off-target serendips} are visually identified as additional brightness peaks in the spatial intensity profile that are offset from the target and spatially coincident with spectral continuum in the two-dimensional spectrum (see Figure~\ref{fig_spec}b).  This includes cases where the spectral continuum is distinct from the target and instances where it is partially blended with the target (i.e. the spatial intensity profile shows a peak with an asymmetric wing).   Once the location of the \textit{off-target serendip} has been visually identified in the spatial intensity profile, an extraction window is manually placed on the two-dimensional spectrum and the one-dimensional spectrum is extracted.  The reduction process then proceeds as outlined in \S\,\ref{sssec_cc} -- \S\,\ref{sssec_quality}: the \textit{zspec} software is run to find the best template match to the one-dimensional spectrum, the LOS velocity is measured, telluric and heliocentric corrections are applied, the LOS velocity error is calculated, and a quality is assigned.  LOS velocities are measured for 244 \textit{off-target serendips}  over the five Keck/DEIMOS multislit masks.

\textit{On-target serendips} are cases where the spatial intensity profile shows only a single peak (i.e. looks like one star) but where 2 sets of spectral absorption features are evident (Ca {\scriptsize\rm{II}}, TiO, etc., see Figure~\ref{fig_spec}c).  The minimum velocity separation for which we detect two distinct superimposed velocities is $\Delta v \sim 50$\kms\  (which is slightly greater than the $\sim 35$\kms\, FWHM of our instrumental resolution near the Ca {\scriptsize\rm{II}} triplet).  We define the fainter of the two superimposed stars as the \textit{on-target serendip}. The best template and LOS velocity match to the \textit{on-target serendip} is determined  during the quality assessment phase.  In cases where the ten best-fit solutions reported by \textit{zspec} include a mix of fits for both the target and the \textit{on-target serendip}, the template and LOS velocity best matching the \textit{on-target serendip} is selected from the list and assigned a quality code ($Q = $3, 4).  In cases where the solutions reported by \textit{zspec} do not show any good matches to the \textit{on-target serendip}, the best-fit is determined by manually shifting stellar templates until a good match is found; these fits are assigned $Q=1$.  Once the LOS velocity has been determined, a heliocentric and an average telluric correction is applied, and the LOS velocity error is calculated.  LOS velocities are measured for 52 \textit{on-target serendips} over the five Keck/DEIMOS multislit masks.

\textit{Off-target superimposed serendips} are a combination of the two categories discussed above.  They consist of cases where continuum that is offset from the primary target turns out to be two perfectly blended \textit{serendips} at the same spatial location.  The one-dimensional spectrum of the \textit{off-target superimposed serendip} is extracted via the method outlined for \textit{off-target serendips}.  The \textit{zspec} software is then run on the one-dimensional spectrum, and analysis proceeds as for \textit{on-target serendips}.  We find 2 such instances of off-target superimposed \textit{serendips} over the five Keck/DEIMOS multislit masks.  

LOS velocities are measured for a total number of 298 \textit{serendips} in all three classes.  The number of \textit{serendip} LOS velocities measured per mask is summarized in Table \ref{tab_obs}.

\subsubsection{Maximum Likelihood Analysis of the Line-of-Sight Velocities}\label{sssec_ml} 

\begin{figure*}[h!]
\plotone{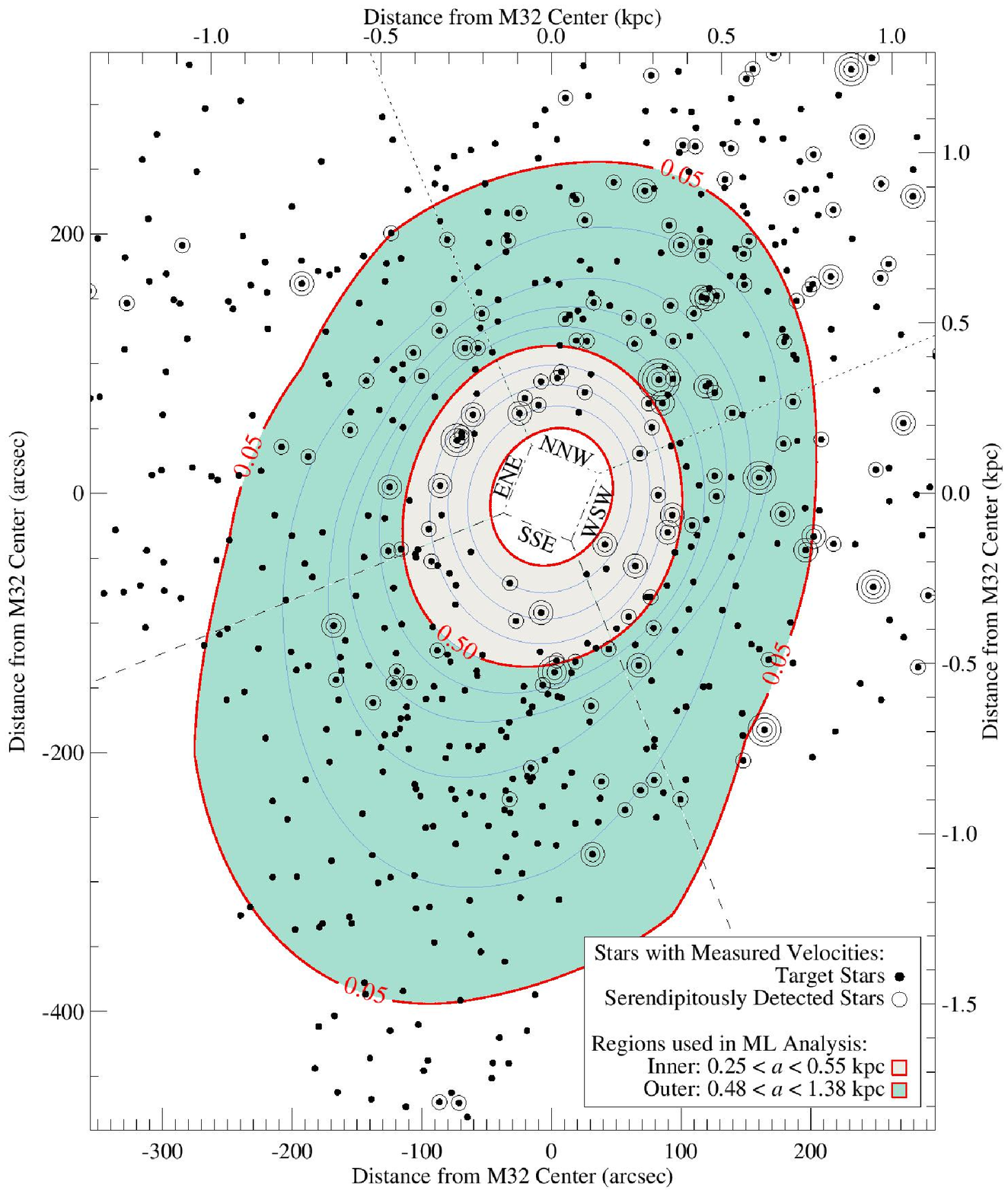}
\caption[A subsample of targets and serendipitous sources (\textit{serendips}) used for kinematical analysis.]{
A subsample of targets (closed circle) and \textit{serendips} (open circle) used for kinematical analysis.  Concentric open circles denote multiple serendipitous detections at a given location.  The contours shown outline M32's predicted fractional $I$-band light contribution relative to M31's inner spheroid and disk.  For the purpose of kinematical analysis, the M32 region is divided into four quadrants: NNW major-axis, SSE major-axis, WSW minor-axis, and ENE minor-axis. In order to determine if strong velocity gradients are present, each quadrant is further divided into three subregions based on the probability, $P$, of M32 membership (derived from two-dimensional surface brightness profiles of M31 and M32).  The subregions for each quadrant are defined as follows: \textit{inner} $\equiv 50\% \leq P \leq 90\%$ (corresponding to a semi-major axis distance of $0.25 \lesssim a \lesssim 0.55$ kpc), and \textit{outer} $\equiv 5\% \leq P < 50\%$ ($0.48 \lesssim a \lesssim 1.38$ kpc).   These subregions are chosen (somewhat arbitrarily) so that the inner and outer subregions contain a reasonable number of M32 stars.  Maximum likelihood fits of Gaussians to the LOS stellar velocities in each of these subregions are shown in Figure~{\ref{fig_ml}}.
}
\label{fig_stars}
\end{figure*}

A subsample of the measured target and \textit{serendip} LOS velocities is selected for further kinematical analysis based on their probability of M32 membership, as determined by sky position.  Stars with a probability $P \ge 5\%$ (see Figure \ref{fig_masks}) are selected as potential M32 candidates.  This reduces the stellar sample from 1,084 to 482 stars and corresponds to a radial range of $0.2 \-- 1.4 \: {\rm kpc}$.  In order to determine if strong velocity and/or velocity dispersion gradients are present along M32's major and minor axes, the subsample of stars is further divided into eight subregions: four quadrants divided into two probability ranges.  The four quadrants, outlined in Figure~\ref{fig_stars}, are centered on M32's axes as follows: north-north-west (NNW) major-axis ($-65^{\circ} \le \phi \le 25^{\circ}$, spanning  $0.2 \-- 0.9 \: {\rm kpc}$), south-south-east (SSE) major-axis ($115^{\circ} \le \phi \le 205^{\circ}$, spanning $0.3 \-- 1.4 \:  {\rm kpc}$), west-south-west (WSW) minor-axis ($205^{\circ} \le \phi \le 295^{\circ}$, spanning $0.2 \-- 1.0 \:  {\rm kpc}$), and east-north-east (ENE) minor-axis ($25^{\circ} \le \phi \le 115^{\circ}$, spanning $0.3 \-- 1.0 \:  {\rm kpc}$).  The lack of symmetry between the contours in M32's SSE and NNW quadrants is due to differing amounts of M31 light contamination; the M31 contamination is significantly worse in M32's NNW region resulting in contours that extend further out on the SSE side of the galaxy.  This effect can also be seen along the minor-axis of the galaxy where the contours extend further out on the ENE side of the galaxy.    The two probability ranges are defined so that each subregion contains a reasonable number of M32 stars: an \textit{inner} region defined by $P \ge 50\%$ ($0.2 \lesssim a \lesssim 0.6 \: {\rm kpc}$), and an \textit{outer} region defined by $ 5\% \le P < 50\%$ ($0.5 \lesssim a \lesssim 1.4 \: {\rm kpc}$).  These eight subregions are illustrated in Figure~\ref{fig_stars}.

We perform maximum likelihood fits of Gaussians to the LOS velocity distribution of stars in the M32 region (all stars with $P \ge 5\%$) and in each of the eight subregions.  While the true shape of the velocity structural components in these regions may differ from pure Gaussians, the use of such models seems appropriate given their ability to characterize the mean velocity and velocity dispersion, the small number of velocity points being assessed, and the absence of any definite physical model.  The individual stellar velocity errors are not included in the maximum likelihood analyses, since these errors ($\sim$4 \kms) are much smaller than the velocity dispersion of any structural component in any of these regions, and therefore contribute insignificantly to the maximum likelihood errors.  The Gaussian fits to the M32-like resolved stellar LOS velocities in each region and corresponding 68\% and 90\% confidence limit errors are summarized in Table \ref{tab_gpoints} and discussed below in turn.

\begin{deluxetable*}{cccccccccccc}
\tabletypesize{\scriptsize}
\tablecaption{Maximum Likelihood Gaussian Best Fit Parameters to M32-like Resolved Stellar Velocities}
\tablewidth{0 pt}
\tablehead{
\multicolumn{1}{c}{} &
\multicolumn{1}{c}{$\bar{r}$ \tablenotemark{a}} &
\multicolumn{1}{c}{RMS($r$) \tablenotemark{b}}&
\multicolumn{1}{c}{$v$ \tablenotemark{c}}&
\multicolumn{2}{c}{$\Delta v$ \tablenotemark{d}}&
\multicolumn{1}{c}{$\sigma$ \tablenotemark{e}}&
\multicolumn{2}{c}{$\Delta\sigma$ \tablenotemark{f}} &
\multicolumn{1}{c}{$N_{\rm{M32}}/N_{\rm{tot}}$ \tablenotemark{g}} &
\multicolumn{2}	{c}{$\Delta N_{\rm{M32}}/N_{\rm{tot}}$ \tablenotemark{h}} \\
\cline{5-6}\cline{8-9}\cline{11-12} 				
\multicolumn{4}{l}{\tiny{Confidence Limits:}} &
\multicolumn{1}{c}{\tiny{(68\%)}}&
\multicolumn{1}{c}{\tiny{(90\%)}} &
\multicolumn{1}{c}{}&
\multicolumn{1}{c}{\tiny{(68\%)}} &
\multicolumn{1}{c}{\tiny{(90\%)}} &
\multicolumn{1}{c}{} &
\multicolumn{1}{c}{\tiny{(68\%)}} &
\multicolumn{1}{c}{\tiny{(90\%)}} 	}
\startdata
\multicolumn{12}{l}{}\\
\multicolumn{12}{l}{\textbf{M32, All Quadrants, Inner + Outer}} \\
	& $174\farcs2$ & $65\farcs0$
	& $-196.9$ & $\pm 3.0$ & $^{+5.0}_{-4.9}$ 
	& $29.9$ & $^{+3.1}_{-2.9}$ & $^{+5.2}_{-4.6}$ 
	& 0.352 &  $\pm 0.028 $ &  $^{+0.047}_{-0.046}$ \\ \\
	
\multicolumn{3}{l}{\small\sc{Major Axis:}}\\
\cline{1-3} \\
\multicolumn{12}{l}{\textbf{M32, NNW Quadrant, Inner}} \\
	& $-95\farcs4$ & $11\farcs9$
	& $-198.7$ & $^{+6.4}_{-6.5}$ &$^{+10.8}_{-11.4}$ 
	& $24.2$ & $^{+6.0}_{-4.4}$ &$^{+11.8}_{-7.0}$ 
	&0.742 &  $^{+0.092}_{-0.110}$ & $^{+0.140}_{-0.189}$\\ \\
\multicolumn{12}{l}{\textbf{M32, NNW Quadrant, Outer}} \\
	& $-201\farcs2$ & $43\farcs1$
	& $-226.4$ & $^{+11.0}_{-11.4}$ &$^{+18.4}_{-20.0}$ 
	& $34.1$ & $^{+9.9}_{-7.4}$ &$^{+19.3}_{-11.8}$ 
	&0.196 &  $^{+0.055}_{-0.052}$ & $^{+0.094}_{-0.084}$\\ \\				
\multicolumn{12}{l}{\textbf{M32, SSE Quadrant, Inner}} \\
	& $123\farcs8$ & $23\farcs4$
	& $-186.3$ & $\pm 9.0$ &$^{+15.2}_{-15.4}$
	& $31.2$ & $^{+7.2}_{-5.5}$ &$^{+13.5}_{-8.5}$
	& 0.663 &  $^{+0.107}_{-0.121}$ &  $^{+0.165}_{-0.204}$\\ \\	
\multicolumn{12}{l}{\textbf{M32, SSE Quadrant, Outer}} \\
	& $225\farcs6$ & $53\farcs5$
	& $-191.0$ & $^{+7.2}_{-7.1}$ &$\pm 11.9$
	& $36.2$ & $^{+6.2}_{-5.4}$ &$^{+11.0}_{-8.8}$
	& 0.311 &  $\pm 0.051$ &  $^{+0.080}_{-0.084}$\\ \\	
	
\multicolumn{3}{l}{\small\sc{Minor Axis:}}\\	
\cline{1-3} \\
\multicolumn{12}{l}{\textbf{M32, ENE Quadrant, Inner}} \\
	& $-106\farcs2$ & $5\farcs1$
	& $-184.4$ & $^{+7.0}_{-8.4}$ &$^{+11.7}_{-14.3}$
	& $26.2$ & $^{+7.8}_{-5.2}$ &$^{+13.5}_{-7.6}$
	& 0.788 &  $^{+0.094}_{-0.113}$&  $^{+0.139}_{-0.195}$\\ \\	
\multicolumn{12}{l}{\textbf{M32, ENE Quadrant, Outer}} \\
	& $-142\farcs6$ & $30\farcs2$
	& $-204.6$ & $^{+6.9}_{-7.0}$ &$^{+11.4}_{-11.8}$
	& $26.0$ & $^{+6.3}_{-5.9}$ &$^{+11.0}_{-9.0}$
	& 0.401 &  $\pm 0.080$&  $^{+0.131}_{-0.130}$\\ \\
\multicolumn{12}{l}{\textbf{M32, WSW Quadrant, Inner}} \\
	& $119\farcs3$ & $21\farcs4$
	& $-194.6$ & $\pm 4.2$ &$^{+7.1}_{-7.0}$
	& $18.1$ & $^{+3.3}_{-2.7}$ &$^{+6.0}_{-4.3}$
	& 0.724 &  $^{+0.083}_{-0.096}$&  $^{+0.128}_{-0.163}$\\ \\
\multicolumn{12}{l}{\textbf{M32, WSW Quadrant, Outer}} \\ 
	& $156\farcs3$ & $28\farcs9$
	& $-179.2$ & $^{+11.7}_{-11.3}$ &$^{+20.2}_{-23.0}$
	& $39.0$ & $^{+9.7}_{-7.8}$ &$^{+18.7}_{-21.8}$
	& 0.297 &  $^{+0.072}_{-0.070}$&  $^{+0.120}_{-0.113}$\\ 		
\enddata
\tablecomments{
Results of the maximum likelihood Gaussian fits to the LOS velocities of the M32-like resolved stellar data (includes targets and \textit{serendips}) in various quadrants located within M32's 5\% predicted fractional light contribution contour (see Figure~\ref{fig_stars}).\\
\tablenotetext{a}{Median projected semi-major/minor axis distance from M32's center for the M32-like stellar population (i.e. stars with velocities in the range $v\pm\sigma$) based on \citet{cho02} \textit{I}-band photometry.  The full M32 region (all quadrants) and major axis quadrants (NNW and SSE) list the semi-major axis distance $a$.  Minor axis quadrants  (ENE and WSW) list the semi-minor axis distance $qa$.  }
\tablenotetext{b}{Root-mean-square of the semi-major/minor axis distance.}
\tablenotetext{c}{Best-fit heliocentric LOS velocity.}
\tablenotetext{d}{Error in best-fit heliocentric LOS velocity ($68\%$ and $90\%$ confidence limits).} 
\tablenotetext{e}{Best-fit velocity dispersion.}
\tablenotetext{f}{Error in best-fit velocity dispersion ($68\%$ and $90\%$ confidence limits).}
\tablenotetext{g}{Best-fit fraction of M32 stars based on kinematics.}
\tablenotetext{h}{Error in best-fit fraction of M32 stars in the sample ($68\%$ and $90\%$ confidence limits).}}
\label{tab_gpoints}
\end{deluxetable*}

\begin{figure}\scalebox{0.42}{\includegraphics[trim=0 15 0 0]{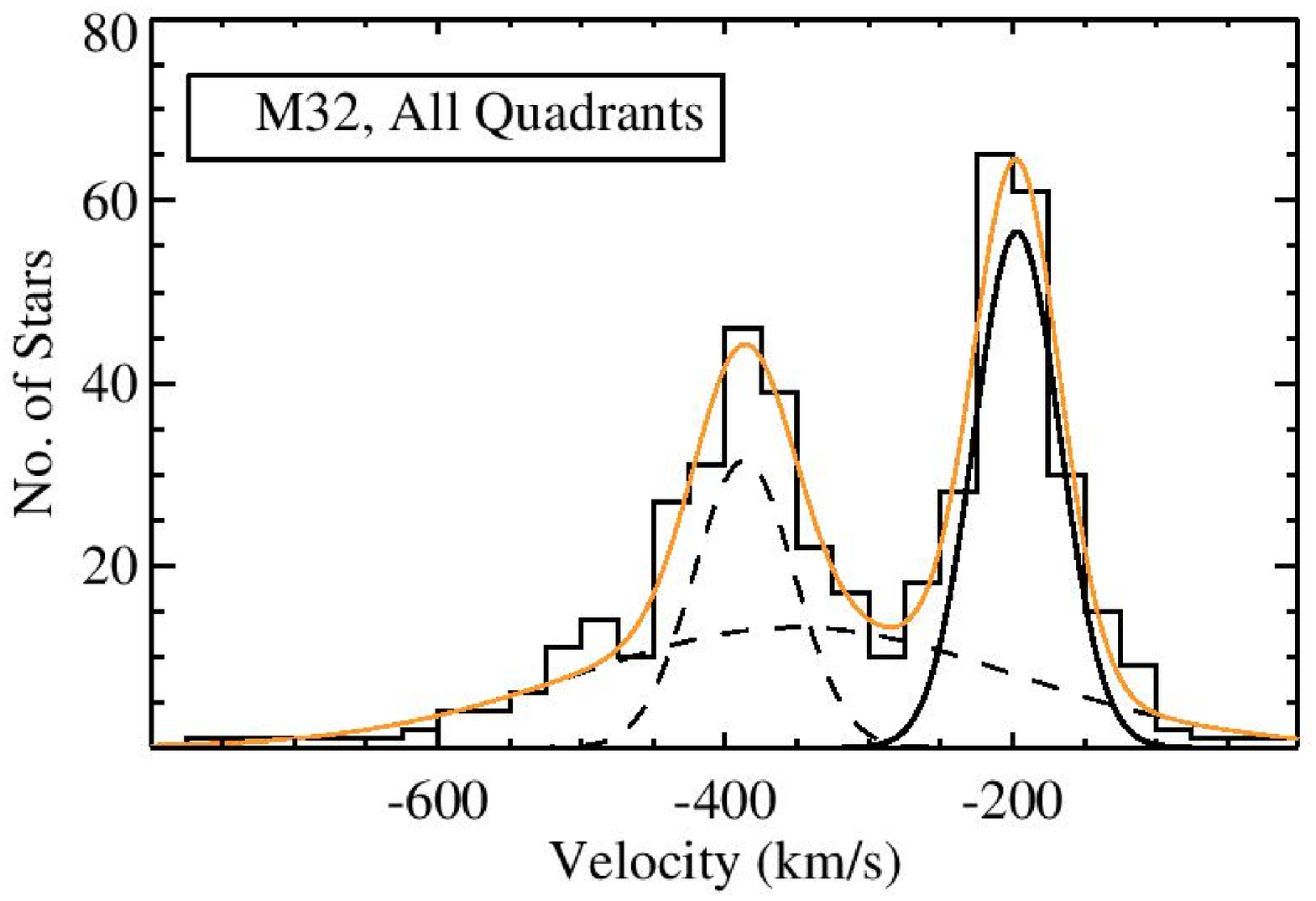}}
\caption[Velocity histogram and maximum likelihood Gaussian fits to the line-of-sight velocities of resolved stars located within the M32 region.]{
Velocity histogram and maximum likelihood Gaussian fits to the LOS velocities of resolved stars located within M32's 5\% predicted fraction light contribution contour (see Figure~\ref{fig_stars}).  The solid black curve shows the fit that is consistent with M32's systemic velocity \citep[$v^{\rm{M32}}_{\rm{sys}}=-200$\kms,][]{fal99}.  The dashed black curves have mean velocities and velocity dispersions consistent with a cold rotating component (M31's disk), with a hot non-rotating component (M31's inner spheroid).  The sum of the Gaussian fits is shown as a solid orange curve.  The fits to the M31-like stellar population are held fixed during the analysis of the quadrants surrounding M32.  
}
\label{fig_ml}
\end{figure}

Figure~\ref{fig_ml} shows maximum likelihood fits of sums of Gaussians to the LOS velocity distribution of stars for the entire M32 region (all stars with $P \ge 5\%$). The observed LOS velocity distribution is well fit by the sum of three Gaussians (red curve).  The narrow Gaussian centered at $v = -196.9 \pm 3.0$\kms\ with a width of $\sigma = 29.9 \pm 2.9$\kms\ (solid black curve) represents $35.2^{+2.8}_{-3.6}\%$ of the stars in the region and is consistent with the systemic velocity of M32 \citep[$v^{\rm{M32}}_{\rm{sys}}=-200$\kms,][]{fal99}.  The two additional populations seen in this region are well fit by a broad Gaussian centered at $v=-350.9$ \kms\, with a width of $\sigma=153.0$ \kms\, representing $41.9\%$ of the stars, and a narrow Gaussian centered at $v=-386.9$ \kms\, with a width of $\sigma = 35.0$ \kms\, representing $22.9\%$ of the stars; these two additional populations are consistent  with the mean velocity of M31's inner spheroid $v^{\rm{M31}}_{\rm{sys}} \sim -300$\kms\ \citep{gil07} and M31's disk, $v^{\rm{M31}}_{\rm{disk}}\sim-400$\kms, respectively \citep{dor12}.

\begin{figure*}\plotone{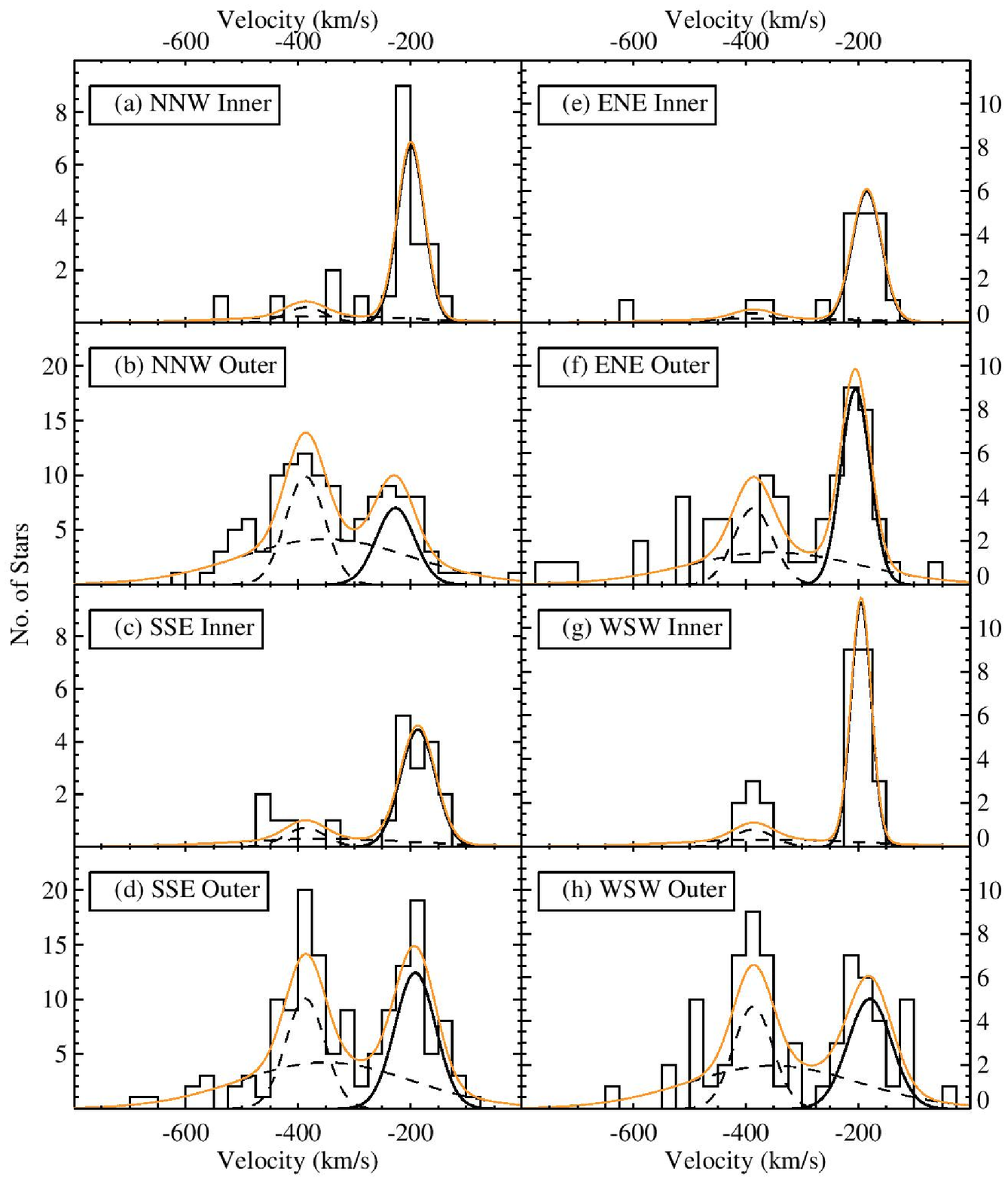}
\caption[Velocity histograms and maximum likelihood Gaussian fits to the line-of-sight velocities of resolved stars along the major axis of M32.]{
Velocity histograms and maximum likelihood Gaussian fits to the LOS velocities of resolved stars along the major and minor axes of M32.
Solid black curves show fits that are consistent with M32's systemic velocity \citep[$v^{\rm{M32}}_{\rm{sys}}=-200$\kms,][]{fal99}.  
Dashed black curves (M31 components) have fixed ratios, velocities and velocity dispersions that are based on a global fit to the M32 region (Figure \ref{fig_ml}).
The sum of the Gaussian fits is shown as a solid orange curve.  Panels a--h show fits to the subregions outlined in Figure \ref{fig_stars}:
\textbf{(a)} the inner region along M32's NNW semi-major axis,
\textbf{(b)} the outer region along M32's NNW semi-major axis,
\textbf{(c)} like (a) but for SSE major axis,
\textbf{(d)} like (b) but for SSE major axis.
\textbf{(e)}  like (a) but for ENE semi-minor axis,
\textbf{(f)}   like (b) but for ENE semi-minor axis,
\textbf{(g)} like (a) but for the WSW minor axis,
\textbf{(h)} like (b) but for the WSW minor axis.
}
\label{fig_mla}
\end{figure*}

Figure~\ref{fig_mla} shows maximum likelihood fits of sums of Gaussians to the LOS velocity distribution of stars to each subregion located along M32's major-axis and minor-axis, respectively.  In each subregion, the two Gaussians that represent M31's inner spheroid and disk (determined in the global fit to the M32 region) are held fixed while the best fit Gaussian parameters to M32 members are searched.  The mean velocity $v$, velocity dispersion $\sigma$, and fraction $N_{\rm{M32}}/N_{\rm{tot}}$ of M32 stars and the corresponding $68\%$ and $90\%$ confidence limits for each subregion are listed in Table~\ref{tab_gpoints}.

We compare the M32/M31 fraction {\it predicted\/} by our 2D surface brightness models
for the two galaxies to the fraction {\it measured\/} from the fit to the LOS
velocity distribution in each subregion.  While the two sets of fractions
follow similar trends across the different subregions, the surface brightness
based predictions of the M32 fraction tend to be systematically lower than the
fractions measured from the LOS velocity distribution analysis (but by only
$\sim1\sigma$ on average).  In any case, this slight discrepancy does not affect
any of the results of this paper.  We only use the surface brightness
based fraction predictions to define the boundaries of the subregions in
which to carry out the kinematical fits, and these boundaries are fairly
arbitrarily defined in any case.  Nevertheless, we consider some possible explanations
for this discrepancy between predicted and measured fractions.
First, M31's disk surface brightness is not uniform
across the M32 region and may well depart from our idealized 2D surface
brightness model.  Second, our spectroscopic target selection tends to bias
the kinematical sample against high surface brightness/crowded patches in
M31's disk, resulting in a bias towards higher M32 fractions.
Third, the translation from integrated $V$-band surface brightness to RGB
star count surface density is likely not the same for M32, M31's inner spheroid, and
M31's disk, and may well vary with radius within the two galaxies.  Finally, there
are uncertainties in our 2D surface brightness models for the two galaxies
associated with $I$- to $V$-band conversion of the surface brightness measurements.

Our kinematically based measurement of the M31 contamination fraction is
relevant for M32 stellar population studies.  For example, in the recent
\citet{mon11} analysis of M32's star-formation history from deep HST ACS/HRC
CMDs, contamination by M31 stars is statistically accounted for using a
control field whose location was chosen on the basis of the shape/orientation
of M31's disk isophotes.  Our kinematical analysis provides an independent
measurement of the M31 contamination fraction in the region of their study,
albeit averaged over a larger area than the narrow HST ACS/HRC field.

  \subsection{Integrated Light Spectroscopy}\label{ssec_integrated}

We begin this subsection by describing observations, and end with measurements of the line-of-sight velocity distribution (LOSVD) along M32's major and minor axes from the integrated light data.  
This subsection is outlined as follows.
In \S\,\ref{sssec_long}, we provide the observing details. 
In \S\,\ref{sssec_longslit}, we summarize the data reduction process.
In \S\,\ref{sssec_kinlong}, we make velocity, velocity dispersion and higher order Gauss-Hermite moment measurements along M32's major and minor axes.

\subsubsection{Observations}\label{sssec_long}

Eight longslit exposures centered on M32 were obtained with Keck/DEIMOS between November 2007-2008.  Six of the longslits were aligned with M32's major-axis ($\phi=160^{\circ}$), and two 
were aligned with M32's minor-axis ($\phi=70^{\circ}$).  Each longslit mask is $\sim 16\arcmin$ long and designed with a series of 4 slits separated by small bridges in order to insure the structural integrity of the mask design.  The observing setup for the longslit exposures is identical to that used for the multislit mask observations discussed in \S\,\ref{sssec_multi}, except that  slit widths of $0\farcs8$ and $1\farcs0$ were used.    The average seeing for these observations was $1\farcs0$, yielding an average spectral resolution of $3.9$ pixels $= 1.3$ \AA.  The observing details for the longslit observations are summarized in Table \ref{tab_long}.

  \begin{deluxetable*}{lllccccc}
\tabletypesize{\scriptsize}
\tablecaption{Keck/DEIMOS Long-Slit Exposures}
\tablewidth{0 pt}
\tablehead{
\colhead{Mask} &
\colhead{Observation} &
\colhead{$\alpha$} &
\colhead{$\delta$} &
\colhead{P.A.} &
\colhead{t$_{\rm{exp}}$} &
\colhead{Seeing} &
\colhead{Slit} \\
\colhead{Name} &
\colhead{Date} &
\colhead{(J2000.0)} &
\colhead{(J2000.0)} &
\colhead{(deg)} &
\colhead{(m)} &
\colhead{FWHM} &
\colhead{Width}
}
\startdata
	Major\_1  & 2007 Nov 14 & 00 42 41.87 & +40 51 57.2  & 160 & 3 & 
	$0\farcs5$ &1\farcs0\\
	Major\_2  & 2008 Nov 14 & 00 42 41.87 & +40 51 57.2  & 160 & 3 & 
	$0\farcs5$ &1\farcs0\\
	Major\_3  & 2008 Oct 01& 00 42 41.87 & +40 51 57.2  & 160 & 5 
	& $\sim1\farcs0$ &1\farcs0\\
	Major\_4  & 2008 Oct 01 & 00 42 41.87 & +40 51 57.2  & 160 & 5 
	& $\sim1\farcs0$ &1\farcs0\\
	Major\_5 & 2008 Nov 24 & 00 42 41.87 & +40 51 57.2  
	& $-$20 & 5 & $1\farcs3$ &0\farcs8\\
	Major\_6  & 2008 Nov 24 & 00 42 41.87 & +40 51 57.2  
	& $-$20 & 5 & $1\farcs3$ &0\farcs8\\
\\
	Minor\_1  & 2008 Nov 24 & 00 42 41.87 & +40 51 57.2  & 70 & 5 &
	$1\farcs3$ &0\farcs8\\ 	Minor\_2  & 2008 Nov 24 & 00 42 41.87 & +40 51 57.2  & 70 & 5 &
	$1\farcs3$ &0\farcs8\enddata
\tablecomments{The units of right ascension ($\alpha$) are in hours, minutes and seconds.  The units of declination ($\delta$) are in degrees, arcminutes and seconds.
}
\label{tab_long}
\end{deluxetable*} 
 
\subsubsection{Data Reduction}\label{sssec_longslit}
 
Processing of the eight Keck/DEIMOS longslit masks from rectification through wavelength calibration is identical to the reduction procedure used for the Keck/DEIMOS multislit masks: the two-dimensional spectra are rectified, flat-field corrected, fringe corrected, and wavelength calibrated using the \textit{spec2d} reduction pipeline (see \S\,\ref{sssec_mreduce}).   
 
One-dimensional spectra are extracted at increasing spatial intervals from the center of M32 along the two-dimensional spectrum.  The location of M32's center is determined by fitting a Moffat profile $+$ 1st order polynomial to the intensity profile of M32 (obtained by collapsing the two-dimensional spectrum in the wavelength direction); the fractional spatial pixel location that corresponds to the peak of the intensity profile fit is defined as M32's center.  This information is used to convert the pixel positions along the two-dimensional spectrum into a distance from M32.  One-dimensional spectra are then extracted using boxcar extraction windows with widths ranging from $1\arcsec$ to $5\arcsec$ (the actual size of the extraction window is determined by the signal-to-noise S/N of the region being extracted), and the Poisson errors are calculated.  The one-dimensional spectra are re-binned logarithmically in the wavelength direction into bins with 13.8\kms/pixel.

\begin{figure}\scalebox{0.55}{\includegraphics[trim=0 0 0 0]{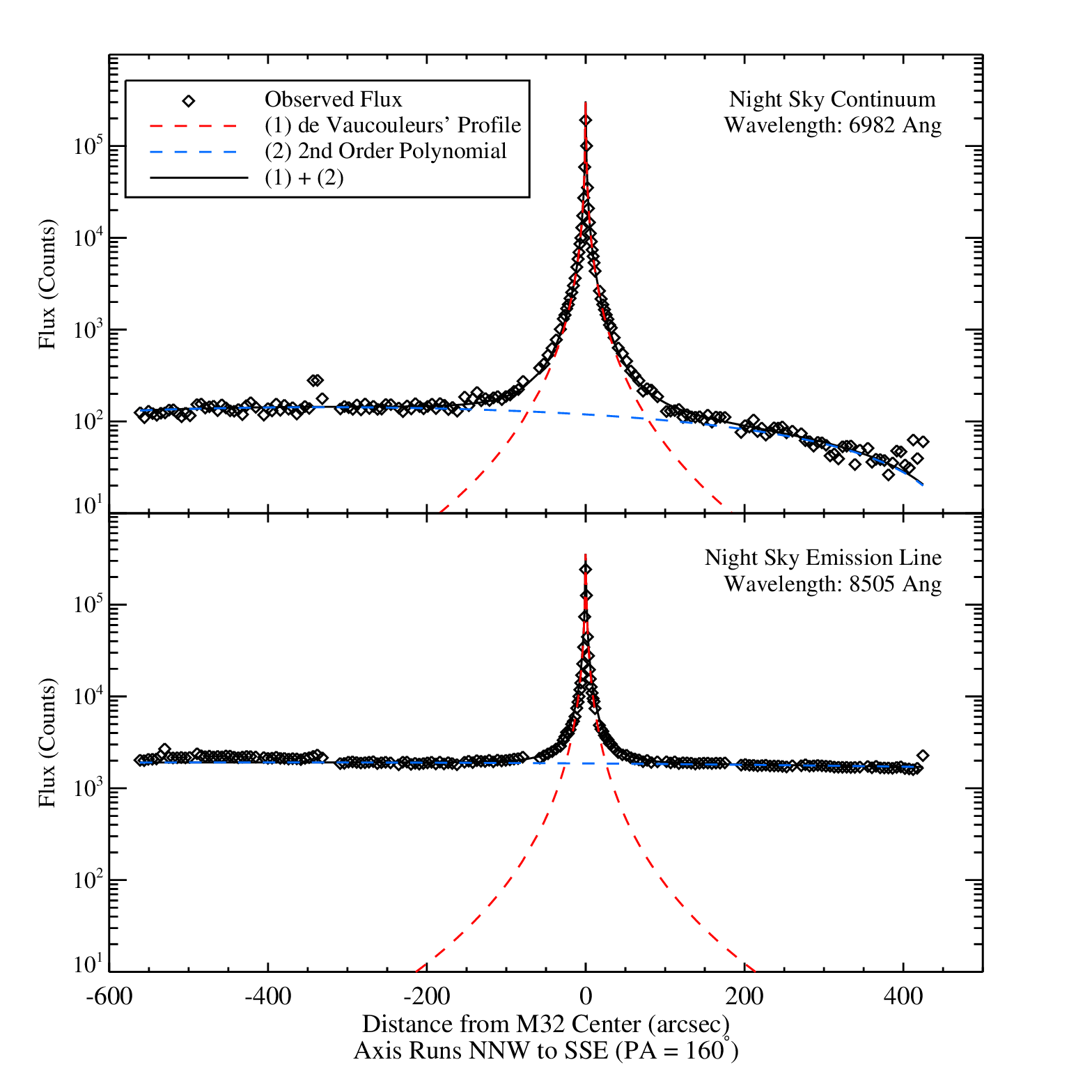}}
\caption[Examples of sky subtraction and the night sky variations in M32's integrated light profile.]{An example of sky subtraction and the night sky variations in M32's integrated light profile.  The plot shows 
light intensity (black diamonds) versus distance from M32's center in two wavelength bins.  The red dashed line provides a fit to M32's light by normalizing a de Vaucouleurs' profile, with $r^{\rm{eff}}_{I} =29\arcsec$ \citep{cho02}, to the observed intensities.  The contaminating light from M31, atmospheric air glow and corrections to the DEIMOS spectrograph response function is measured by subtracting off the normalized de Vaucouleurs' profile from the observed intensities and fitting a 2nd order polynomial to the residual intensities beyond $90\arcsec$ from M32's center (dashed blue line).   The sum of the two profiles is shown as a solid black line.  \textbf{Top:} Wavelength bin corresponding to night sky continuum.  The gradient seen in the continuum (dashed blue line) is a result of light contamination from M31. \textbf{Bottom:} Wavelength bin corresponding to a peak of a night sky emission line.   The M31 gradient is not visible against the dominant night sky emission.
}
\label{fig_devauc}
\end{figure}

Next, sky subtraction is performed on the one-dimensional spectra.  Extreme care is taken to properly subtract the light from all contaminating sources, which includes M31, atmospheric air glow, and imperfect flat-fielding (resulting from differences in the illumination between the internal flat-field exposure and the on sky science exposure).  Sky subtraction is performed separately for each wavelength bin.  First, the intensity profile as a function of distance from M32's center is obtained for each wavelength bin.  A normalized de Vaucouleurs' profile with $r^{\rm{eff}}_{I}  = 29\arcsec$ is fit to and subtracted from each intensity profile outside the inner 5\arcsec of M32's center, where excessive flux is present \citep{ken87, cho02}; this produces a ``M32-free" intensity profile for each wavelength bin.  Next, a 2nd order polynomial representing all the contaminating light sources is fit to each "M32-free" intensity profile at positions beyond $90\arcsec$ from M32's center.  The 2nd order polynomial fits are subtracted from the original intensity profiles, and then reassembled into one-dimensional sky-subtracted spectra.  An example of the fitting and subtraction process is shown in Figure~\ref{fig_devauc}.

Once sky subtraction is complete the major-axis spectra observed with the $0\farcs8$ slit width (see Table \ref{tab_long}) are Gaussian smoothed in the spectral direction in order to match the spectral resolution of the $1\farcs0$ slit width spectra (since the minor-axis science spectra were all observed with the same slit width they do not require smoothing for co-addition).  The smoothing length $\sigma_{\rm smooth}$ needed for the $0\farcs8$ major-axis spectra is determined to be 0.42\,\AA\ using the following formula:
\begin{equation}
\sigma_{\rm smooth} = {\rm m}\frac{\Delta\lambda}{\Delta x}\: \sqrt{\frac{w_{2}^{2} - w_{1}^{2}}{8\ln{2}}}
\label{eqn_smooth} 
\end{equation}
where $w_{2}=1\farcs0$ is the desired slit width, $w_{1}=0\farcs8$ is the observed slit width, ${\rm m} =0.606$ is the anamorphic demagnification factor at the central wavelength 7800\,\AA,  $\Delta\lambda = 0.32$\,\AA/pixel is the spectral dispersion, and $\Delta x=0\farcs1185$/pixel is the spatial scale.  

Next, the one-dimensional sky-subtracted spectra from individual exposures
and original fine spatial bins are coadded, with cosmic ray rejection, into
broader spatial bins.  These broader bins are designed to achieve a minimum
spectral S/N ratio per pixel of 25 averaged over the Ca {\scriptsize\rm{II}}
triplet region.  This S/N threshold limits the radial extent of our
kinematical analysis to $\leq 90\arcsec$ and $\leq 60\arcsec$ on the major
and minor axes, respectively.  About 15\% of the individual spectra are
excluded from the coadds as they were deemed, upon visual inspection, to
suffer from bad subtraction of night sky emission lines and/or other
systematic errors (e.g., slit edges and bad columns).  Inclusion of these bad
spectra in the coadds would lead to choppier, less symmetric kinematical
profiles.  No coaddition is performed on spectra located at
$|r_{\rm{M32}}|\le2\arcsec$ as the individual spectra exceed the S/N
threshold in these bright inner regions.

     \subsubsection{Measurement of Velocity, Velocity Dispersion and Higher Order Gauss-Hermite Moments}\label{sssec_kinlong}

\begin{figure*}\plotone{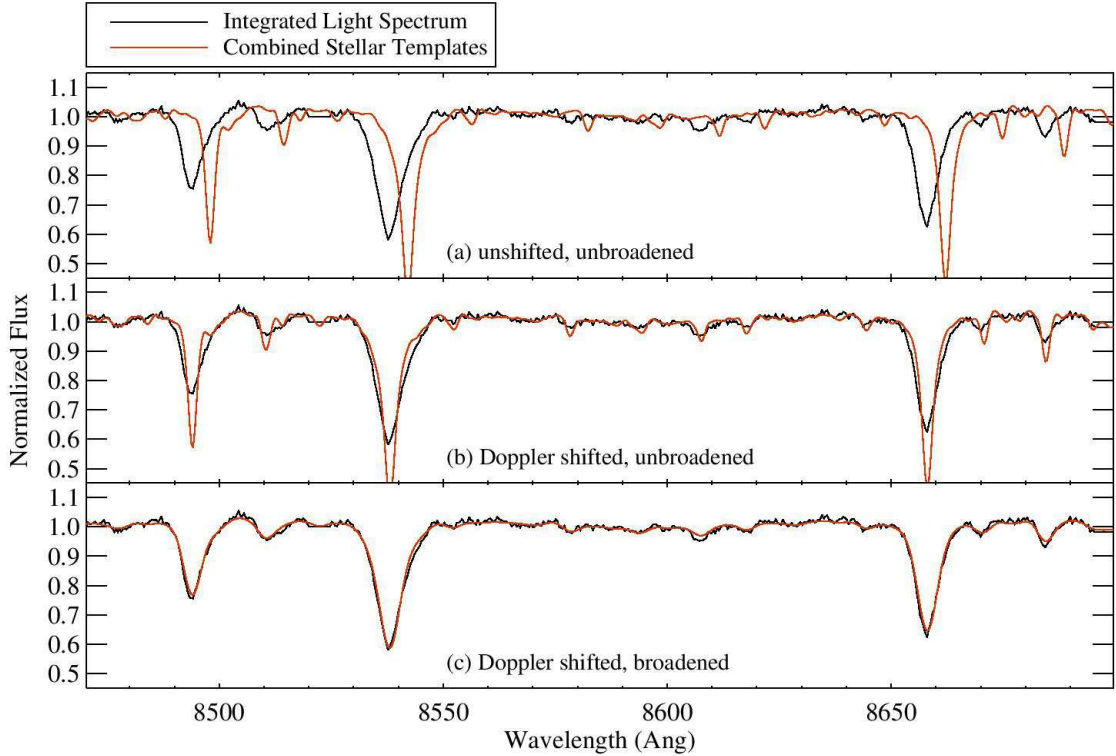}
\caption[Illustration of the fitting of a composite stellar spectral templateto an integrated light one-dimensional spectrum of M32.]{
An illustration of the fitting of a composite stellar spectral template (red) to a $1\arcsec$ (4 pc) wide integrated light one-dimensional spectrum of M32.  The composite template is constructed by optimally combining 16 stellar templates using a genetic algorithm.  
The three panels illustrate the steps used to measure the LOSVD of the M32 spectrum .  
(a) Observed M32 spectrum and an un-broadened composite stellar template (with no Doppler shift applied).  
(b) The Doppler velocity of the M32 spectrum is measured by shifting the un-broadened composite stellar template to match the features of the M32 spectrum. 
(c) The velocity dispersion of the M32 spectrum is determined by broadening the composite stellar template so that it matches the observed M32 spectrum.
} 
\label{fig_ga}
\end{figure*}

The mean velocity $v$, velocity dispersion $\sigma$, and Gauss-Hermite moments $h3$ through $h6$ 
that best fit the observed LOSVD are determined using the \textit{pixfit} software developed and described by \cite{van94}.  Briefly, the software determines the best-fit absorption line strength parameter ($\gamma$), LOS velocity ($v$), and velocity dispersion ($\sigma$) for each one-dimensional integrated light science spectrum by minimizing $\chi^2$ between scaled, Doppler-shifted, Gaussian-broadened spectral templates and the science spectra (see Figure~\ref{fig_ga}).  Symmetric and anti-symmetric deviations from Gaussianity are then measured by expanding the fits to include higher order Gauss-Hermite moments  \citep[$h3$ through $h6$,][]{van93} for spectra with an average S/N per pixel $\gtrsim 40$.
 
The rms uncertainties in our best-fit parameters $x = [v$, $\sigma$, $h3$, $h4$, $h5$, $h6$] are determined using the Poisson error in the flux of the input science spectrum.  Since the Poisson-based error estimates $\Delta^{P}x$ are bound to be underestimates and do not account for systematic errors such as imperfect subtraction of night sky emission lines, template mismatch, and residual detector artifacts, we attempt to derive more realistic error estimates, $\Delta x = f \times\Delta^{P}x$, where the error scale factor $f$ is derived empirically by assuming that the true kinematical profiles in the higher S/N outer regions of M32 ($|r| > 5\arcsec$, $S/N \gtrsim 40$) are symmetric and smooth.  The symmetry assumption implies that one half of the anti-symmetric profiles ($v$, $h3$ and $h5$) can be point-reflected onto its other half, and that one half of the symmetric profiles ($\sigma$, $h4$ and $h6$) can be mirror-reflected onto its other half.  The smoothness assumption implies that the folded data points  $x_{\rm data}(r_{i})$ can be compared to a second-order polynomial $x_{\rm poly}(r)$ that has been fit to the data.  Specifically, we require that deviations of the folded profiles from the smooth polynomial has a $\chi^2$ per degree of freedom ($\chi^{2}_{\rm DOF}$) of unity:
\begin{equation}
\chi^{2}_{\rm DOF} \equiv \frac{1}{N}  \sum_{i} \frac{[x_{\rm{data}}(r_{i}) - x_{\rm{poly}}(r_{i})]^{2}}{[\Delta x(r_{i})]^{2}} = 1,
\end{equation}
thereby defining the error scale factor: 
\begin{equation}
f^{2} = \frac{1}{N} \sum_{i}  \frac{[x_{\rm{data}}(r_{i}) - x_{\rm{poly}}(r_{i})]^{2}}{[\Delta^{P} x(r_{i})]^{2}}. 
\end{equation}
Error scale factors are calculated for each the major- and minor-axis velocity, velocity dispersion and combined Gauss-Hermite moment profiles.

 The LOSVD is analyzed over the wavelength range 8470--8700\,\AA\ (i.e., the region around the Ca {\scriptsize\rm{II}} triplet: 8498, 8542, 8662\,\AA).  This wavelength range is selected to maximize the LOSVD ``signal" while minimizing systematic errors.  A comparative analysis of the LOSVD distribution between independent sections of the spectrum (``blue": 6500--8470\,\AA\ and ``red'': 8470--8900\,\AA) indicates that systematic errors are relatively large in the ``blue" portion.  Scaling the Poisson -based errors on the LOSVD parameters to ensure ``blue'' vs.\ ``red'' agreement overpredicts the errors by at least a factor of two relative to those derived from the full spectrum and a scaling based on the assumption of smoothness and symmetry/antisymmetry in the radial LOSVD profiles.

The spectral template used to analyze the LOSVD is constructed by combining weighted stellar templates.  The choice of spectral template is of particular importance as mismatch between the science spectra and spectral template can result in significant systematic errors \citep{rix92}.  Given that M32 is composed of a variety of stellar types, it is not surprising that the galaxy spectra are not well fit by any single stellar spectrum.  For this reason, a composite stellar template is constructed using a weighted linear superposition of stellar templates (details on the stellar templates can be found in \S\,\ref{sssec_cc}).  Since the stellar templates were observed with $0\farcs7$ wide slits, the templates are smoothed in the spectral direction so that they match the resolution of the science spectra; the templates are smoothed by $\sigma_{\rm smooth} = 0.50$\,\AA\, for comparison to the $1\farcs0$ resolution major-axis data, and by $\sigma_{\rm smooth} = 0.27$\,\AA\, for comparison to the $0\farcs8$ resolution minor-axis data (see Equation \ref{eqn_smooth}).  The optimal weights for the stellar templates are determined using a genetic algorithm (GA) as configured by \citet{how08}.  The GA locates the global minimum in the weight parameter space by minimizing the $\chi^2$ between a high S/N science spectrum and co-added weighted stellar templates, which have been Doppler shifted to match the science spectrum using cross-correlation and Gaussian smoothed using an initial velocity dispersion estimate.  
We assume that the velocity dispersions across the different stellar components are the same (i.e. not a function of age, stellar mass or metallicity) so that the co-added stellar templates are all smoothed with the same amount of velocity dispersion.
Because template matching becomes increasingly difficult with lower S/N spectra, the procedure is run on a single high S/N spectrum ($r_{\rm{M32}}=-2\farcs0$, S/N $=134$) with a well measured velocity dispersion \citep[$\sigma=56.2$\kms,][]{van94b}.  The intention is to use an unshifted, un-broadened version of the weighted composite stellar template to measure the velocity and velocity dispersion profiles for each science spectrum, thereby treating the abundance and population gradients in M32 as roughly constant with radius such that only the LOSVD, signal and noise of each spectrum varies.  Once the optimal combination of stellar templates is found, the continuum of the composite stellar template is adjusted to better match the science spectrum.  This is achieved by fitting a 5th order polynomial to the ratio of (science spectrum)/(broadened composite stellar template), both to the blue and red sides of the spectrum separately, and multiplying the resulting polynomial by the un-broadened template to remove any low order frequency differences.  The final match between the science spectrum and composite stellar template is shown in Figures~{\ref{fig_ga}(c--d)}.

Table \ref{tab_vpoints} lists the best-fit parameters and scaled errors to the LOSVD as a function of radius.
The error scale factors applied to the major-axis Poisson-based error estimates are $f_{v} = 3.8$, $f_{\sigma} = 2.8$, and
$f_{h} = 1.9$.
The error scale factors applied to the minor-axis Poisson-based error estimates are $f_{v} = 2.1$, $f_{\sigma} = 1.9$, and
$f_{h} = 2.1$.

\clearpage
\LongTables
\begin{deluxetable*}{*{15}{r}c}\tabletypesize{\scriptsize}
\tablecaption{M32 Integrated Light Profiles for Velocity, Velocity Dispersion, and Gauss-Hermite Moments}
\tablewidth{0 pt}
\tablehead{
\colhead{$r$\tablenotemark{a}} &
\colhead{$dr$\tablenotemark{b}} &
\colhead{$v$\tablenotemark{c}} &
\colhead{$\Delta v$\tablenotemark{d}} &
\colhead{$\sigma$\tablenotemark{e}} &
\colhead{$ \Delta \sigma$\tablenotemark{f}} &
\colhead{$h_3$\tablenotemark{g}} &
\colhead{$\Delta h_3$\tablenotemark{h}} &
\colhead{$h_4$\tablenotemark{g}} &
\colhead{$\Delta h_4$\tablenotemark{h}} &
\colhead{$h_5$\tablenotemark{g}} &
\colhead{$ \Delta h_5$\tablenotemark{h}} &
\colhead{$h_6$\tablenotemark{g}} &
\colhead{$\Delta h_6$\tablenotemark{h}} &
\colhead{S/N\tablenotemark{i}} &
\colhead{Mask Name\tablenotemark{j}}
}
\startdata
\multicolumn{16}{l}{\textbf{M32 Major Axis ($\phi=160^{\circ}$), Integrated Light}} \\
$  -85\farcs0$&$  10\farcs0$&$ -23.6$&$   7.3$&$  40.0$&$   7.9$&\multicolumn{1}{c}{\--} &\multicolumn{1}{c}{\--} &\multicolumn{1}{c}{\--} &\multicolumn{1}{c}{\--} &\multicolumn{1}{c}{\--} &\multicolumn{1}{c}{\--} &\multicolumn{1}{c}{\--} &\multicolumn{1}{c}{\--} &$        28$&\multicolumn{1}{c}{\--}\\
$  -70\farcs0$&$   5\farcs0$&$ -19.5$&$   6.0$&$  40.6$&$   6.9$&\multicolumn{1}{c}{\--} &\multicolumn{1}{c}{\--} &\multicolumn{1}{c}{\--} &\multicolumn{1}{c}{\--} &\multicolumn{1}{c}{\--} &\multicolumn{1}{c}{\--} &\multicolumn{1}{c}{\--} &\multicolumn{1}{c}{\--} &$        33$&\multicolumn{1}{c}{\--}\\
$  -60\farcs0$&$   5\farcs0$&$ -17.6$&$   5.0$&$  47.9$&$   5.1$&$ 0.033$&$ 0.085$&$ 0.102$&$ 0.102$&$-0.002$&$ 0.092$&$-0.103$&$ 0.101$&$        47$&\multicolumn{1}{c}{\--}\\
$  -50\farcs0$&$   5\farcs0$&$ -19.2$&$   2.5$&$  41.1$&$   2.8$&$ 0.047$&$ 0.058$&$-0.067$&$ 0.073$&$-0.028$&$ 0.062$&$ 0.081$&$ 0.071$&$        80$&\multicolumn{1}{c}{\--}\\
$  -40\farcs0$&$   5\farcs0$&$ -18.9$&$   2.4$&$  42.6$&$   2.6$&$-0.001$&$ 0.052$&$-0.040$&$ 0.066$&$ 0.053$&$ 0.057$&$ 0.038$&$ 0.064$&$        81$&\multicolumn{1}{c}{\--}\\
$  -32\farcs5$&$   2\farcs5$&$ -18.1$&$   2.1$&$  48.7$&$   2.4$&$ 0.053$&$ 0.037$&$-0.019$&$ 0.045$&$-0.019$&$ 0.041$&$ 0.029$&$ 0.045$&$        97$&\multicolumn{1}{c}{\--}\\
$  -27\farcs5$&$   2\farcs5$&$ -23.6$&$   1.3$&$  47.7$&$   1.4$&$ 0.055$&$ 0.024$&$ 0.001$&$ 0.029$&$-0.005$&$ 0.026$&$ 0.017$&$ 0.029$&$       151$&\multicolumn{1}{c}{\--}\\
$  -22\farcs5$&$   2\farcs5$&$ -24.2$&$   1.4$&$  52.5$&$   1.5$&$ 0.054$&$ 0.021$&$ 0.011$&$ 0.026$&$-0.002$&$ 0.024$&$-0.002$&$ 0.025$&$       149$&\multicolumn{1}{c}{\--}\\
$  -17\farcs5$&$   2\farcs5$&$ -30.1$&$   0.9$&$  51.8$&$   1.0$&$ 0.048$&$ 0.014$&$ 0.008$&$ 0.017$&$ 0.002$&$ 0.016$&$ 0.010$&$ 0.017$&$       222$&\multicolumn{1}{c}{\--}\\
$  -12\farcs5$&$   2\farcs5$&$ -36.2$&$   0.8$&$  54.4$&$   0.9$&$ 0.048$&$ 0.012$&$ 0.017$&$ 0.014$&$ 0.022$&$ 0.013$&$ 0.001$&$ 0.014$&$       250$&\multicolumn{1}{c}{\--}\\
$   -7\farcs5$&$   2\farcs5$&$ -43.8$&$   0.6$&$  55.7$&$   0.6$&$ 0.050$&$ 0.008$&$ 0.016$&$ 0.009$&$ 0.019$&$ 0.009$&$ 0.002$&$ 0.009$&$       360$&\multicolumn{1}{c}{\--}\\
$   -4\farcs5$&$   0\farcs5$&$ -44.9$&$   0.9$&$  58.8$&$   1.0$&$ 0.060$&$ 0.011$&$ 0.030$&$ 0.013$&$0.000$&$ 0.012$&$-0.015$&$ 0.013$&$       240$&\multicolumn{1}{c}{\--}\\
$   -3\farcs5$&$   0\farcs5$&$ -44.9$&$   0.7$&$  59.7$&$   0.8$&$ 0.059$&$ 0.009$&$ 0.014$&$ 0.011$&$ 0.003$&$ 0.010$&$ 0.001$&$ 0.011$&$       287$&\multicolumn{1}{c}{\--}\\
$   -2\farcs5$&$   0\farcs5$&$ -45.7$&$   0.8$&$  60.8$&$   0.9$&$ 0.058$&$ 0.009$&$ 0.024$&$ 0.011$&$ 0.006$&$ 0.011$&$-0.009$&$ 0.012$&$       258$&\multicolumn{1}{c}{\--}\\
$   -1\farcs9$&$   0\farcs6$&$ -43.4$&$   1.1$&$  66.2$&$   1.2$&$ 0.066$&$ 0.011$&$ 0.039$&$ 0.014$&$-0.009$&$ 0.013$&$-0.026$&$ 0.014$&$       199$&Major\_6\\
$   -1\farcs9$&$   0\farcs6$&$ -44.9$&$   1.1$&$  62.8$&$   1.2$&$ 0.067$&$ 0.012$&$ 0.034$&$ 0.014$&$-0.009$&$ 0.014$&$-0.019$&$ 0.014$&$       199$&Major\_5\\
$   -1\farcs3$&$   0\farcs6$&$ -45.0$&$   0.9$&$  62.0$&$   1.0$&$ 0.064$&$ 0.010$&$ 0.030$&$ 0.013$&$-0.006$&$ 0.012$&$-0.015$&$ 0.013$&$       228$&Major\_3\\
$   -1\farcs3$&$   0\farcs6$&$ -45.3$&$   0.9$&$  62.2$&$   1.0$&$ 0.067$&$ 0.010$&$ 0.028$&$ 0.012$&$-0.004$&$ 0.012$&$-0.010$&$ 0.012$&$       236$&Major\_4\\
$-   0\farcs9$&$   0\farcs6$&$ -42.8$&$   1.0$&$  70.1$&$   1.0$&$ 0.069$&$ 0.009$&$ 0.014$&$ 0.011$&$ 0.002$&$ 0.011$&$-0.003$&$ 0.011$&$       236$&Major\_2\\
$-   0\farcs7$&$   0\farcs6$&$ -22.6$&$   0.9$&$  81.3$&$   0.9$&$ 0.056$&$ 0.006$&$ 0.010$&$ 0.007$&$-0.027$&$ 0.007$&$-0.009$&$ 0.007$&$       312$&Major\_6\\
$-   0\farcs7$&$   0\farcs6$&$ -23.7$&$   0.8$&$  79.1$&$   0.8$&$ 0.062$&$ 0.006$&$ 0.015$&$ 0.007$&$-0.024$&$ 0.007$&$-0.011$&$ 0.007$&$       319$&Major\_5\\
$-   0\farcs1$&$   0\farcs6$&$  -7.5$&$   0.7$&$  87.9$&$   0.7$&$ 0.040$&$ 0.005$&$-0.017$&$ 0.005$&$-0.016$&$ 0.005$&$ 0.014$&$ 0.005$&$       399$&Major\_3\\
$-   0\farcs1$&$   0\farcs6$&$  -5.1$&$   0.7$&$  89.4$&$   0.7$&$ 0.034$&$ 0.004$&$-0.016$&$ 0.005$&$-0.018$&$ 0.005$&$ 0.014$&$ 0.005$&$       415$&Major\_4\\
$    0\farcs3$&$   0\farcs6$&$  14.7$&$   0.9$&$  87.3$&$   0.9$&$-0.054$&$ 0.006$&$-0.012$&$ 0.006$&$ 0.029$&$ 0.007$&$ 0.009$&$ 0.007$&$       322$&Major\_2\\
$    0\farcs5$&$   0\farcs5$&$  22.9$&$   0.8$&$  81.6$&$   0.9$&$-0.047$&$ 0.006$&$ 0.003$&$ 0.007$&$ 0.022$&$ 0.007$&$-0.004$&$ 0.007$&$       320$&Major\_6\\
$    0\farcs5$&$   0\farcs5$&$  23.8$&$   0.8$&$  79.3$&$   0.8$&$-0.056$&$ 0.006$&$-0.001$&$ 0.007$&$ 0.027$&$ 0.007$&$ 0.002$&$ 0.007$&$       326$&Major\_5\\
$    1\farcs0$&$   0\farcs6$&$  44.3$&$   0.8$&$  66.2$&$   0.8$&$-0.097$&$ 0.008$&$ 0.022$&$ 0.010$&$ 0.048$&$ 0.009$&$-0.012$&$ 0.010$&$       280$&Major\_3\\
$    1\farcs1$&$   0\farcs6$&$  46.2$&$   0.8$&$  66.2$&$   0.8$&$-0.087$&$ 0.008$&$ 0.007$&$ 0.010$&$ 0.038$&$ 0.009$&$ 0.005$&$ 0.010$&$       278$&Major\_4\\
$    1\farcs0$&$   0\farcs6$&$  46.6$&$   1.2$&$  61.9$&$   1.3$&$-0.097$&$ 0.014$&$ 0.014$&$ 0.017$&$ 0.049$&$ 0.016$&$-0.001$&$ 0.017$&$       176$&Major\_2\\
$    1\farcs7$&$   0\farcs6$&$  43.5$&$   1.1$&$  66.5$&$   1.1$&$-0.081$&$ 0.011$&$ 0.016$&$ 0.013$&$ 0.035$&$ 0.013$&$-0.012$&$ 0.013$&$       204$&Major\_6\\
$    1\farcs7$&$   0\farcs6$&$  44.3$&$   1.1$&$  64.6$&$   1.1$&$-0.103$&$ 0.011$&$ 0.026$&$ 0.014$&$ 0.056$&$ 0.013$&$-0.020$&$ 0.014$&$       204$&Major\_5\\
$    2\farcs5$&$   0\farcs5$&$  46.3$&$   0.6$&$  61.3$&$   0.7$&$-0.091$&$ 0.007$&$ 0.021$&$ 0.009$&$ 0.038$&$ 0.008$&$-0.010$&$ 0.009$&$       337$&\multicolumn{1}{c}{\--}\\
$    3\farcs5$&$   0\farcs5$&$  45.3$&$   1.0$&$  59.6$&$   1.1$&$-0.092$&$ 0.012$&$ 0.005$&$ 0.015$&$ 0.041$&$ 0.014$&$ 0.005$&$ 0.015$&$       207$&\multicolumn{1}{c}{\--}\\
$    4\farcs5$&$   0\farcs5$&$  44.7$&$   0.9$&$  58.2$&$   1.0$&$-0.087$&$ 0.011$&$ 0.010$&$ 0.014$&$ 0.030$&$ 0.013$&$ 0.004$&$ 0.014$&$       229$&\multicolumn{1}{c}{\--}\\
$    7\farcs5$&$   2\farcs5$&$  42.7$&$   0.5$&$  56.2$&$   0.6$&$-0.089$&$ 0.007$&$ 0.013$&$ 0.009$&$ 0.032$&$ 0.008$&$0.000$&$ 0.009$&$       384$&\multicolumn{1}{c}{\--}\\
$   12\farcs5$&$   2\farcs5$&$  33.0$&$   0.8$&$  56.5$&$   0.9$&$-0.076$&$ 0.011$&$ 0.014$&$ 0.013$&$ 0.028$&$ 0.013$&$-0.002$&$ 0.013$&$       250$&\multicolumn{1}{c}{\--}\\
$   17\farcs5$&$   2\farcs5$&$  28.1$&$   1.0$&$  54.1$&$   1.1$&$-0.090$&$ 0.015$&$ 0.016$&$ 0.018$&$ 0.053$&$ 0.016$&$-0.009$&$ 0.018$&$       205$&\multicolumn{1}{c}{\--}\\
$   22\farcs5$&$   2\farcs5$&$  23.4$&$   1.3$&$  51.6$&$   1.4$&$-0.077$&$ 0.021$&$ 0.015$&$ 0.025$&$ 0.049$&$ 0.023$&$0.000$&$ 0.025$&$       155$&\multicolumn{1}{c}{\--}\\
$   27\farcs5$&$   2\farcs5$&$  21.9$&$   1.7$&$  53.4$&$   1.8$&$-0.058$&$ 0.024$&$ 0.006$&$ 0.030$&$ 0.037$&$ 0.027$&$-0.004$&$ 0.029$&$       128$&\multicolumn{1}{c}{\--}\\
$   32\farcs5$&$   2\farcs5$&$  22.0$&$   5.5$&$  46.5$&$   5.9$&$-0.099$&$ 0.102$&$-0.013$&$ 0.125$&$ 0.032$&$ 0.112$&$ 0.019$&$ 0.123$&$        36$&\multicolumn{1}{c}{\--}\\
$   40\farcs0$&$   5\farcs0$&$  16.2$&$   2.2$&$  47.6$&$   2.3$&$-0.068$&$ 0.039$&$-0.026$&$ 0.048$&$ 0.039$&$ 0.043$&$ 0.028$&$ 0.047$&$        95$&\multicolumn{1}{c}{\--}\\
$   50\farcs0$&$   5\farcs0$&$  16.9$&$   3.4$&$  43.6$&$   3.7$&$-0.105$&$ 0.070$&$-0.020$&$ 0.086$&$ 0.108$&$ 0.075$&$ 0.034$&$ 0.084$&$        63$&\multicolumn{1}{c}{\--}\\
$   60\farcs0$&$   5\farcs0$&$  15.8$&$   4.4$&$  45.6$&$   4.9$&$-0.046$&$ 0.083$&$-0.074$&$ 0.101$&$ 0.006$&$ 0.089$&$ 0.105$&$ 0.099$&$        52$&\multicolumn{1}{c}{\--}\\
$   70\farcs0$&$   5\farcs0$&$  14.7$&$   5.9$&$  42.4$&$   6.4$&\multicolumn{1}{c}{\--} &\multicolumn{1}{c}{\--} &\multicolumn{1}{c}{\--} &\multicolumn{1}{c}{\--} &\multicolumn{1}{c}{\--} &\multicolumn{1}{c}{\--} &\multicolumn{1}{c}{\--} &\multicolumn{1}{c}{\--} &$        39$&\multicolumn{1}{c}{\--}\\
$   80\farcs0$&$   5\farcs0$&$  18.7$&$   7.5$&$  29.0$&$   9.8$&\multicolumn{1}{c}{\--} &\multicolumn{1}{c}{\--} &\multicolumn{1}{c}{\--} &\multicolumn{1}{c}{\--} &\multicolumn{1}{c}{\--} &\multicolumn{1}{c}{\--} &\multicolumn{1}{c}{\--} &\multicolumn{1}{c}{\--} &$        25$&\multicolumn{1}{c}{\--}\\
\\
\multicolumn{16}{l}{\textbf{M32 Minor Axis ($\phi=250^{\circ}$), Integrated Light}} \\
$ -40\farcs0$&$   5\farcs0$&$  -3.7$&$   2.0$&$  41.0$&$   2.5$&$-0.119$&$ 0.077$&$-0.119$&$ 0.094$&$ 0.068$&$ 0.083$&$ 0.087$&$ 0.093$&$        67$&\multicolumn{1}{c}{\--}\\
$ -32\farcs5$&$   2\farcs5$&$  -2.1$&$   2.0$&$  39.3$&$   2.3$&$-0.189$&$ 0.074$&$-0.189$&$ 0.092$&$ 0.168$&$ 0.080$&$ 0.418$&$ 0.091$&$        63$&\multicolumn{1}{c}{\--}\\
$ -27\farcs5$&$   2\farcs5$&$  -0.9$&$   1.8$&$  49.4$&$   2.2$&$-0.061$&$ 0.050$&$-0.061$&$ 0.060$&$ 0.033$&$ 0.056$&$ 0.051$&$ 0.060$&$        69$&\multicolumn{1}{c}{\--}\\
$ -22\farcs5$&$   2\farcs5$&$   0.9$&$   1.5$&$  51.1$&$   1.9$&$-0.017$&$ 0.042$&$-0.017$&$ 0.049$&$ 0.030$&$ 0.046$&$ 0.021$&$ 0.050$&$        79$&\multicolumn{1}{c}{\--}\\
$ -17\farcs5$&$   2\farcs5$&$  -0.2$&$   1.0$&$  52.8$&$   1.3$&$-0.034$&$ 0.027$&$-0.034$&$ 0.032$&$ 0.027$&$ 0.030$&$-0.008$&$ 0.032$&$       118$&\multicolumn{1}{c}{\--}\\
$ -12\farcs5$&$   2\farcs5$&$  -0.8$&$   0.9$&$  55.9$&$   1.2$&$-0.035$&$ 0.023$&$-0.035$&$ 0.027$&$ 0.013$&$ 0.025$&$ 0.051$&$ 0.027$&$       129$&\multicolumn{1}{c}{\--}\\
$  -7\farcs5$&$   2\farcs5$&$   0.5$&$   0.6$&$  60.2$&$   0.8$&$ 0.005$&$ 0.013$&$ 0.005$&$ 0.016$&$ 0.003$&$ 0.015$&$-0.008$&$ 0.016$&$       193$&\multicolumn{1}{c}{\--}\\
$  -4\farcs5$&$   0\farcs5$&$   3.0$&$   0.9$&$  61.9$&$   1.2$&$-0.003$&$ 0.019$&$-0.003$&$ 0.023$&$0.000$&$ 0.022$&$-0.033$&$ 0.023$&$       128$&\multicolumn{1}{c}{\--}\\
$  -3\farcs5$&$   0\farcs5$&$   3.4$&$   0.8$&$  65.2$&$   1.0$&$-0.009$&$ 0.015$&$-0.009$&$ 0.018$&$ 0.008$&$ 0.018$&$-0.012$&$ 0.018$&$       156$&\multicolumn{1}{c}{\--}\\
$  -2\farcs5$&$   0\farcs5$&$   4.6$&$   0.6$&$  69.0$&$   0.8$&$0.000$&$ 0.011$&$0.000$&$ 0.012$&$ 0.006$&$ 0.012$&$-0.022$&$ 0.013$&$       211$&\multicolumn{1}{c}{\--}\\
$  -1\farcs0$&$   0\farcs6$&$   9.3$&$   0.6$&$  79.9$&$   0.7$&$-0.006$&$ 0.008$&$-0.006$&$ 0.010$&$ 0.002$&$ 0.010$&$-0.005$&$ 0.010$&$       241$&Minor\_3\\
$  -1\farcs0$&$   0\farcs6$&$  10.4$&$   0.6$&$  78.9$&$   0.7$&$-0.021$&$ 0.008$&$-0.021$&$ 0.009$&$ 0.004$&$ 0.010$&$-0.003$&$ 0.010$&$       247$&Minor\_4\\
$   0\farcs1$&$   0\farcs6$&$  11.0$&$   0.5$&$  87.4$&$   0.6$&$-0.035$&$ 0.006$&$-0.035$&$ 0.007$&$ 0.013$&$ 0.007$&$ 0.016$&$ 0.007$&$       323$&Minor\_3\\
$   0\farcs2$&$   0\farcs6$&$  12.0$&$   0.5$&$  88.4$&$   0.6$&$-0.032$&$ 0.006$&$-0.032$&$ 0.006$&$ 0.010$&$ 0.006$&$ 0.016$&$ 0.007$&$       342$&Minor\_4\\
$   1\farcs3$&$   0\farcs6$&$   8.1$&$   0.6$&$  77.4$&$   0.8$&$-0.020$&$ 0.009$&$-0.020$&$ 0.011$&$ 0.009$&$ 0.011$&$-0.006$&$ 0.011$&$       222$&Minor\_3\\
$   1\farcs4$&$   0\farcs6$&$   7.6$&$   0.6$&$  76.5$&$   0.8$&$-0.017$&$ 0.009$&$-0.017$&$ 0.011$&$ 0.012$&$ 0.011$&$-0.002$&$ 0.011$&$       219$&Minor\_4\\
$   2\farcs5$&$   0\farcs5$&$   3.6$&$   0.6$&$  68.0$&$   0.8$&$-0.002$&$ 0.011$&$-0.002$&$ 0.013$&$ 0.002$&$ 0.013$&$-0.021$&$ 0.014$&$       196$&\multicolumn{1}{c}{\--}\\
$   3\farcs5$&$   0\farcs5$&$   1.7$&$   0.8$&$  64.8$&$   1.0$&$-0.012$&$ 0.016$&$-0.012$&$ 0.018$&$ 0.018$&$ 0.018$&$-0.028$&$ 0.018$&$       151$&\multicolumn{1}{c}{\--}\\
$   4\farcs5$&$   0\farcs5$&$   1.2$&$   1.0$&$  63.5$&$   1.3$&$ 0.009$&$ 0.019$&$ 0.009$&$ 0.023$&$-0.001$&$ 0.022$&$-0.031$&$ 0.023$&$       125$&\multicolumn{1}{c}{\--}\\
$   7\farcs5$&$   2\farcs5$&$   0.2$&$   0.6$&$  59.7$&$   0.8$&$ 0.010$&$ 0.014$&$ 0.010$&$ 0.016$&$ 0.000$&$ 0.016$&$-0.011$&$ 0.016$&$       187$&\multicolumn{1}{c}{\--}\\
$  12\farcs5$&$   2\farcs5$&$  -0.3$&$   0.9$&$  55.7$&$   1.2$&$-0.009$&$ 0.022$&$-0.009$&$ 0.026$&$-0.002$&$ 0.025$&$-0.023$&$ 0.026$&$       127$&\multicolumn{1}{c}{\--}\\
$  17\farcs5$&$   2\farcs5$&$  -1.6$&$   1.0$&$  51.4$&$   1.3$&$0.000$&$ 0.028$&$0.000$&$ 0.033$&$ 0.004$&$ 0.031$&$-0.002$&$ 0.033$&$       113$&\multicolumn{1}{c}{\--}\\
$  22\farcs5$&$   2\farcs5$&$  -5.0$&$   1.4$&$  48.8$&$   1.9$&$-0.012$&$ 0.044$&$-0.012$&$ 0.053$&$ 0.006$&$ 0.049$&$ 0.011$&$ 0.053$&$        79$&\multicolumn{1}{c}{\--}\\
$  27\farcs5$&$   2\farcs5$&$  -4.9$&$   1.8$&$  48.3$&$   2.3$&$-0.082$&$ 0.053$&$-0.082$&$ 0.064$&$ 0.096$&$ 0.059$&$ 0.022$&$ 0.064$&$        67$&\multicolumn{1}{c}{\--}\\
$  32\farcs5$&$   2\farcs5$&$  -4.5$&$   1.9$&$  47.2$&$   2.4$&$-0.059$&$ 0.059$&$-0.059$&$ 0.071$&$ 0.063$&$ 0.065$&$ 0.060$&$ 0.071$&$        64$&\multicolumn{1}{c}{\--}\\
$  40\farcs0$&$   5\farcs0$&$   2.6$&$   2.9$&$  42.3$&$   3.9$&$ 0.022$&$ 0.117$&$ 0.022$&$ 0.142$&$-0.020$&$ 0.126$&$-0.002$&$ 0.141$&$        37$&\multicolumn{1}{c}{\--}\\
$  50\farcs0$&$   5\farcs0$&$   2.1$&$   2.7$&$  41.9$&$   3.6$&$-0.040$&$ 0.107$&$-0.040$&$ 0.131$&$-0.004$&$ 0.114$&$ 0.062$&$ 0.128$&$        42$&\multicolumn{1}{c}{\--}\\
$  60\farcs0$&$   5\farcs0$&$  -2.5$&$   4.0$&$  49.6$&$   5.3$&\multicolumn{1}{c}{\--} &\multicolumn{1}{c}{\--} &\multicolumn{1}{c}{\--} &\multicolumn{1}{c}{\--} &\multicolumn{1}{c}{\--} &\multicolumn{1}{c}{\--} &\multicolumn{1}{c}{\--} &\multicolumn{1}{c}{\--} &$        30$&\multicolumn{1}{c}{\--}\\
$  75\farcs0$&$  10\farcs0$&$  -2.2$&$   3.9$&$  39.0$&$   4.8$&\multicolumn{1}{c}{\--} &\multicolumn{1}{c}{\--} &\multicolumn{1}{c}{\--} &\multicolumn{1}{c}{\--} &\multicolumn{1}{c}{\--} &\multicolumn{1}{c}{\--} &\multicolumn{1}{c}{\--} &\multicolumn{1}{c}{\--} &$        28$&\multicolumn{1}{c}{\--}\\
\enddata
\tablecomments{
The orientation of the axes has been defined as follows: major-axis runs NNW to SSE ($\phi=160^{\circ}$), minor-axis runs ENE to WSW ($\phi=250^{\circ}$).  The systemic velocity of M32 is measured as $-199.7 \pm 0.5$\kms\ along the major-axis and $-200.1 \pm 1.8$\kms\ along the minor-axis. \\
\tablenotetext{a}{Projected distance along the axis in arcseconds}
\tablenotetext{b}{Spatial width of the bin in arcseconds}
\tablenotetext{c}{Measured LOS velocity in \kms }
\tablenotetext{d}{Scaled Poisson error in measured LOS velocity in \kms (where the scale factor $f=3.8$ and $2.1$ for the major- and minor-axis, respectively)}
\tablenotetext{e}{Measured velocity dispersion in \kms}
\tablenotetext{f}{Scaled Poisson error in measured velocity dispersion in \kms (where the scale factor $f=2.8$ and $1.9$ for the major- and minor-axis, respectively)}
\tablenotetext{g}{Gauss-Hermite moments for spectra with S/N $\gtrsim 40$}
\tablenotetext{h}{Scaled Poisson error in the Gauss-Hermite moments (where the scale factor $f=1.9$ and $2.1$ for the major- and minor-axis, respectively}
\tablenotetext{i}{Average S/N per pixel of the spectrum}
\tablenotetext{j}{Mask names for the spectra that were not coadded (i.e. spectra located at $|r_{\rm M32}|\le 2\arcsec$)}
}
\label{tab_vpoints}
\end{deluxetable*}

  \subsection{An Integrated View of M32's Kinematics}\label{ssec_vprofile}
 
\begin{figure*}
\scalebox{0.74}{\includegraphics[trim=0 0 0 0]{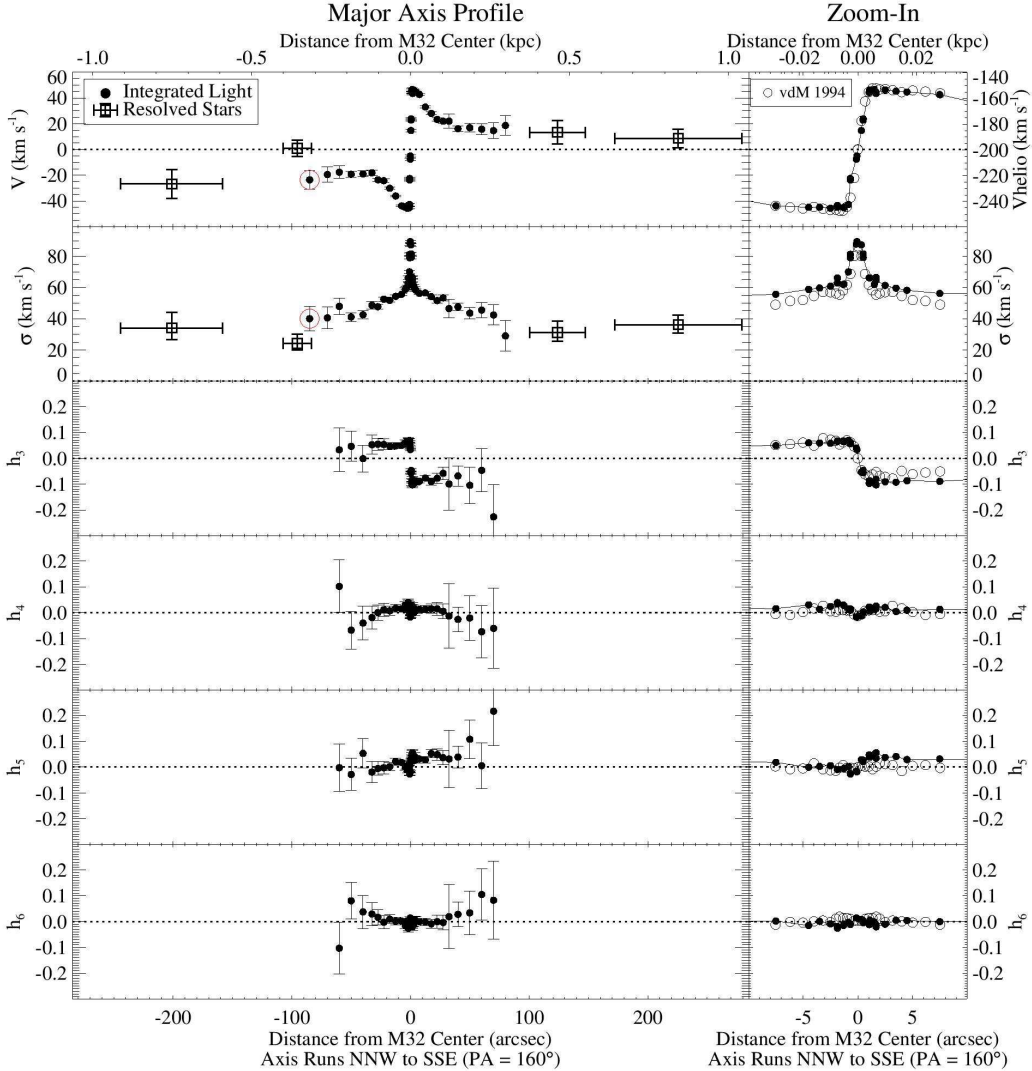}}
\caption[Best fit parameters to M32's major-axis line-of-sight velocity distribution from measurements of integrated light and Gaussian fits to the resolved stellar population.]{
Best fit parameters to M32's major-axis ($\phi=160^{\circ}$) LOSVD from measurements of integrated light (solid circles) and Gaussian fits to the resolved stellar population (open squares). Rows from top to bottom: M32's major-axis mean velocity, velocity dispersion, and the Gauss-Hermite moments $h_3$, $h_4$, $h_5$, and $h_6$ profiles.  Left column: The full radial extent of each profile with $1\sigma$ (68\% confidence limit) error bars.  The integrated light measurement at $-85\farcs0$ (circled in red) is possibly affected by M31 contamination, thereby pulling the line-of-sight velocity point more negative. Right column: Comparison with \citet{van94b} measurements (open circles) shows good agreement.  Error bars are not shown, but are generally smaller than the point size.  The distance scale, assuming a distance to M32 of 785 kpc, is $1\arcsec = 4$pc.
}
\label{fig_majprofile}
\end{figure*}

\begin{figure*}
\scalebox{0.74}{\includegraphics[trim=0 0 0 0]{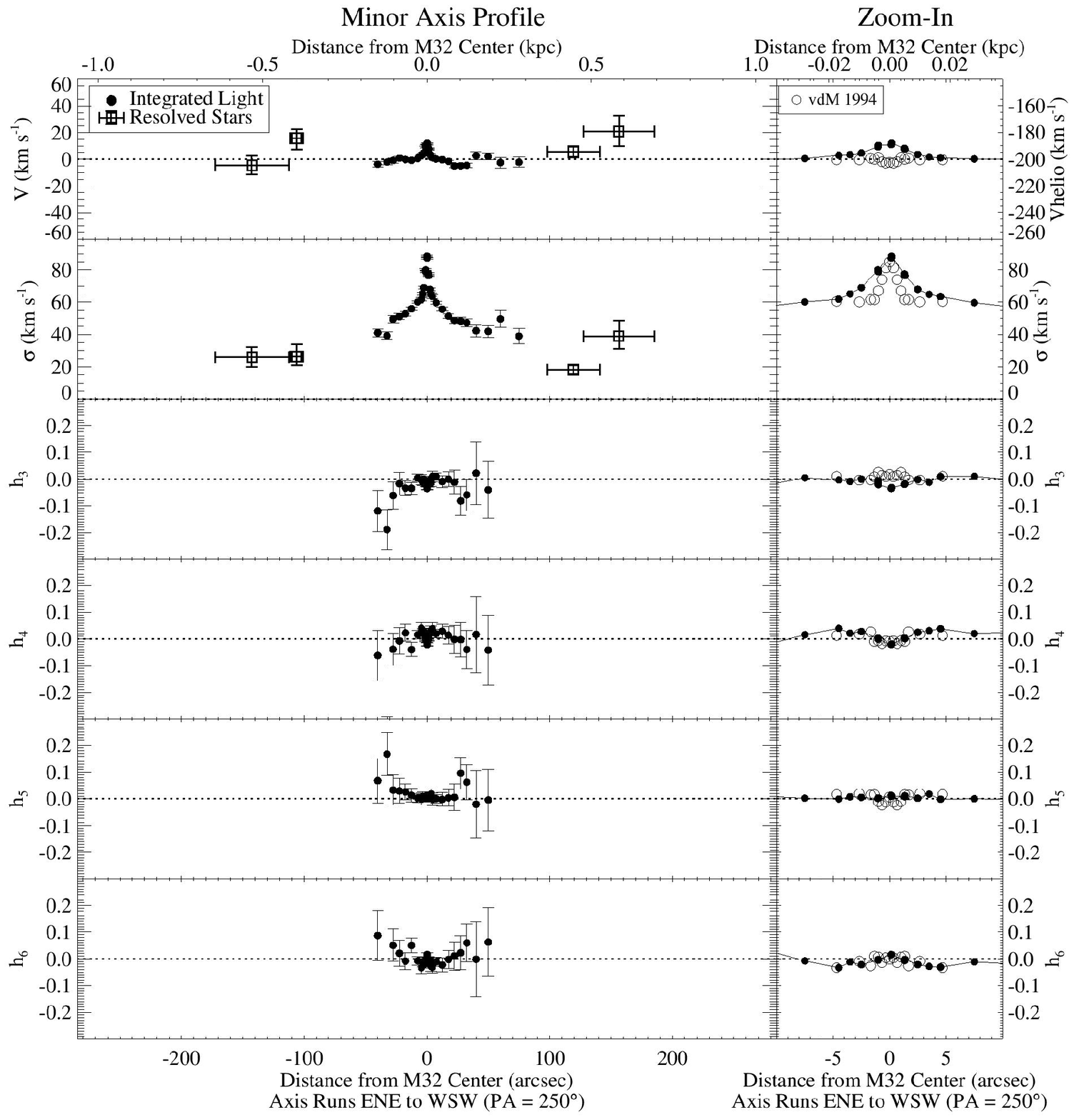}}
\caption[Best fit parameters to M32's minor-axis line-of-sight velocity distribution from measurements of integrated light and Gaussian fits to the resolved stellar population.]{
Same as Figure \ref{fig_majprofile} but for M32's minor-axis ($\phi=250^{\circ}$).  The slight mismatch seen at the center in the mean velocity profile (an apparent peak in our data and an apparent dip in \citet{van94b}), and the $h3$ and $h5$ profiles is an artifact that results from a slight mis-centering of the longslit in our observations and that of \citet{van94b}.
}
\label{fig_minprofile}
\end{figure*}

The results of M32's major- and minor-axis LOSVD analyses are shown in Figures~\ref{fig_majprofile} and \ref{fig_minprofile}, respectively.
Our measurements push out much further that what has been observed to date.
The velocity and velocity dispersion profiles are measured out to $\sim 230\arcsec$ along the major-axis and $\sim 160\arcsec$ along the minor-axis using Gaussian fits to the resolved stellar population, with 
integrated light measurements reaching out to $85\arcsec$ along the major-axis and $75\arcsec$  along the minor-axis. 
 Because the integrated light velocity measurement at $85\arcsec$ on the NNW major-axis of M32 is likely affected by M31 light (pulling the velocity data point more negative), we exclude this data point from our analysis in \S\,\ref{sec_analysis}.
Measurement of the Gauss-Hermite moment profiles using integrated light extends out to $70\arcsec$ along the major-axis and $50\arcsec $ along the minor-axis.  
The Gaussian fits to the resolved stellar population extend well beyond the major-axis photometric distortion radius of $150$\arcsec\ (minor-axis distance of 130\arcsec) where 
surface photometry of M32 shows a sharp upward break in the surface brightness profile and elongation and twisting of the elliptical isophotes, distortions that may be the result of tidal interaction with M31.
Our profiles do not show evidence of sharp kinematical gradients across the distortion region.  

While the emphasis of our study is on the largely uncharted regions of M32, it is instructive to check whether our measurements agree with previous measurements of the inner regions. 
A comparison of our integrated light measurements with \citet{van94b} observations of M32's core is shown in Figures~\ref{fig_majprofile} and \ref{fig_minprofile} (right).  
The profiles show good agreement, with minor differences attributable to spatial resolution, slit position and template matching. 
Previous studies using integrated light to measure the mass of M32's central black hole have resulted in numerous detailed velocity measurements of the core \citep{ton87,dre88,van94b,ver02,van10}.  
The most radially extensive of these studies measures a two-dimensional mean velocity, velocity dispersion, and Gauss-Hermite moments $h3$ and $h4$ out to 30\arcsec\, using wide field SAURON observations \citep{cap07,van10}.  However, due to the large instrumental dispersion of the instrument, measurements for velocity dispersion, $h3$ and $h4$ may not be very accurate \citep{van10}.  
Studies with higher instrumental resolution measure 
a velocity dispersion profile out to 13\arcsec\ along the major-axis and 22\arcsec\ along the minor-axis \citep{dre88,van94b}, and Gauss-Hermite 
moments out to 8\arcsec\ along the major-axis and 22\arcsec\ along the minor-axis \citep{van94b}.   
 
M32's profiles appear smooth and symmetric.
Our best-fit major-axis velocity profile rises steeply to a maximum observed rotational velocity of $v_{\rm max}=46$\kms\, at $r= 1\farcs0$ (4pc).   
Our best-fit minor-axis velocity profile is relatively flat, with the small cusp seen at the center of the profile resulting from a $0\farcs25$ mis-centering of the longslit during observation.  The observed centrally rising velocity dispersion profile peaks at $\sigma_{\rm max} \approx 90$\kms. 
These measurements are consistent with previous ground-based observations.  Observations of the core using a narrower slit with higher spatial resolution result in higher measurements for velocity and velocity dispersion \citep{van97,jos01}.

The systemic velocity of M32 is measured as $-199.7 \pm 0.5$\kms\ along the major-axis and $-200.1 \pm 1.8$\kms\ along the minor-axis.  These measurements are made by assuming point reflection symmetry of the velocity profile at $r\le10\arcsec$ along the major-axis, and outside the central cuspy region along the minor-axis.  These measurements are consistent with the Zwicky Catalog value of $-200 \pm 6$\kms\ \citep{fal99}.

\section{DYNAMICAL MODELS}\label{sec_analysis}

\subsection{Modeling Approach}

To interpret our new kinematical data we fit it with axisymmetric
dynamical equilibrium models. The models are constructed using
Schwarzschild's orbit superposition technique. Over the past decade
such models have become a standard in the field, and they have been
applied and tested repeatedly with different software implementations
on data for the galaxy M32
\citep[e.g.][]{van98b,cre99,ver02,val04,cap06}. \citet{van10} recently
extended the modeling approach to include triaxial
configurations. However, they found that M32 is actually best fit by
models that are close to axisymmetric and edge-on. We therefore
restrict our analysis to such models here.

Datasets of high spatial resolution obtained with the Hubble Space
Telescope \citep{van98a, jos01} have demonstrated that M32 hosts a
massive black hole in its center. Its mass is $M_{\rm BH} = (2.4 \pm
1.0) \times 10^6 \Msun$ \citep[][consistent with determinations by
previous authors]{van10}.  The observational setup for our new Keck
data was optimized for spatial extent, and not spatial
resolution. Rather than treat $M_{\rm BH}$ as a free parameter in
our analysis, we therefore keep it fixed at $(2.4 \pm 1.0) \times 10^6
\Msun$ in all our models.

The data obtained by other authors generally differ from our own only
inside the central few arcsec, due to differences in spatial
resolution. This is important for understanding the black hole mass,
but that is not a focus of the present study. At intermediate radii,
where existing datasets have some overlap with our own, the data are
generally mutually consistent (see, e.g.,
Figure~\ref{fig_majprofile}). Rather than to combine data from
different authors, we therefore model here only the new Keck data,
which span all observationally accessible radii along two axes. The
quality of our fits discussed below, and their consistency with
previous published work, lead us to believe that addition of integral
field data \citep[available for $\lesssim 28\arcsec$ from][]{cap06}
would not significantly alter the main results.

For the modeling we use the software developed and described by
\citet{van98b}, with similar numerical parameter and grid settings as
described therein. The models start with a luminosity density
distribution, with is transformed to a mass density under the
assumption of a constant mass-to-light ratio $M/L$. The luminosity
density is chosen to fit the observed major axis surface brightness
profile. For this we used the same $V$-band brightness profile as in
\citet{van98b}, but we transformed this to the $I$-band using the
known $V-I$ color \citep{lau98, ton01}. Slightly different from
\citet{van98b}, we adopt a foreground extinction $A_B = 0.35$ and
projected axial ratio $q = 0.76$ from \citet{van10}. This makes our
$M/L$ values directly comparable to theirs, since those authors also
adopt the same distance as we do here.

Given this approach, the $M/L$ is the only free parameter to optimize
the fit to the data. We construct models with different $M/L$. The
best-fit model is identified as the one that yields residuals with the
overall minimum $\chi^2$. 
We restrict our calculations here to models with a constant
mass-to-light ratio for several reasons: (1) such models are easiest
to calculate; (2) such models have been found to adequately fit
existing datasets for M32; (3) such models are a useful reference
before considering more detailed modeling; and (4) such models can be
used in many cases to prove the presence of a dark halo (namely, if no
constant mass-to-light ratio model can fit the data).

\subsection{Data-Model Comparison: Integrated-Light}

\begin{figure*}\plotone{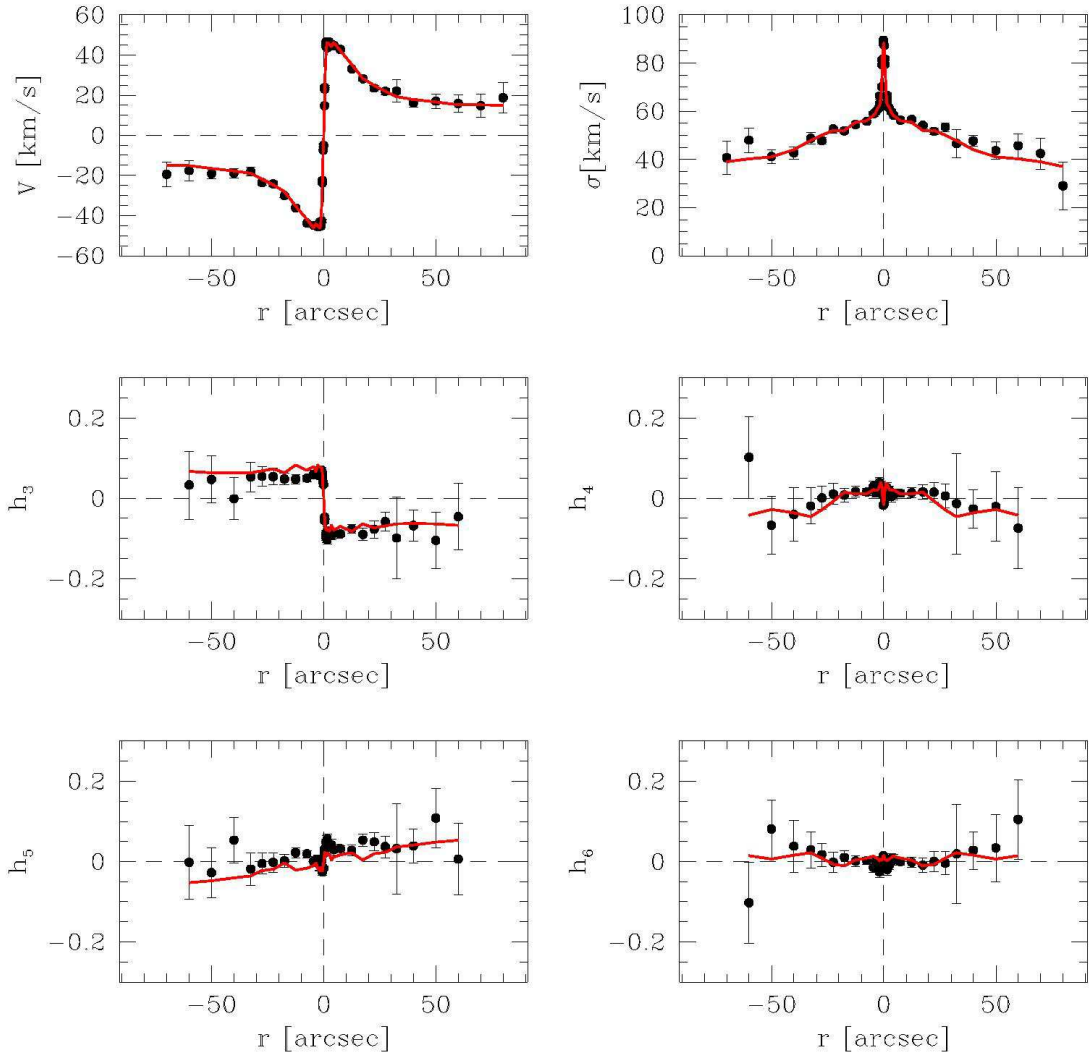}
\caption{Data-model comparison for the long-slit data along the major 
axis. Data points with error bars are shown in black. Model curves shown 
in red display the predictions of the best-fit axisymmetric edge-on model.
A constant mass-to-light ratio model with a central black hole provides an
adequate fit out the the largest radii accessible with integrated light.}
\label{fig_plotmaj2}
\end{figure*}

\begin{figure*}\plotone{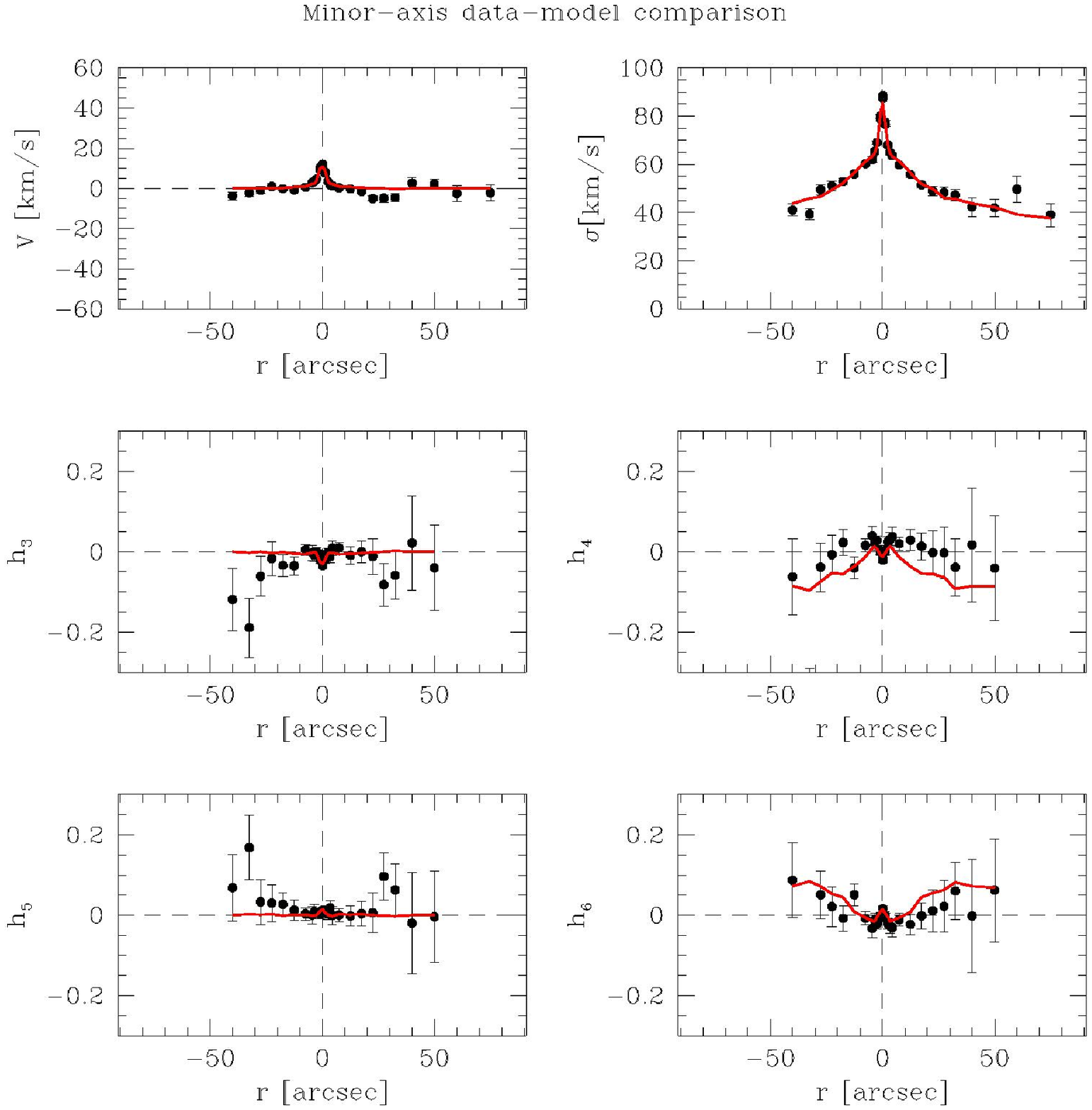}
\caption{Same as Figure~\ref{fig_plotmaj2}, but for the minor axis. The
small central peak in mean velocity $V$ is reproduced by a $0\farcs25$
perpendicular offset of the slit from the galaxy center.}
\label{fig_plotmin2}
\end{figure*}

The predictions of the best-fit model are compared to the long-slit
integrated light measurements in Figures~\ref{fig_plotmaj2} and
\ref{fig_plotmin2}, for the major and minor axis respectively.

The data-model comparison shows that integrated light kinematics for
M32 are well-fit by an axisymmetric constant mass-to-light ratio model
(with a central black hole). While the fit is not perfect
($\chi^2/N_{\rm DF} = 2.35$), all the trends in the data as function
of radius are reasonably well matched by the model.

The finding that a constant mass-to-light ratio model adequately fits
integrated-light kinematical data is consistent with what has been
found by previous authors. However, our work extends this result to a
radius that is three or more times larger than the region assessed by
prior studies. This is a non-trivial finding, since one might have
expected to start seeing the tell-tale signs of a possible dark halo
at $\sim 3 r_{\rm eff}$. But no such signs are readily evident.

The ($I$-band) mass-to-light ratio of our best-fit model is $M/L =
1.24$. \citet{van10} found acceptable (triaxial) models in the range
$M/L = 1.4 \pm 0.2$. So while our best-fit value is lower than
preferred by those authors, it is within the allowed range. The fact
that, if anything, our fits over a much larger radial range yield a
{\it lower} $M/L$ is important. If in reality the velocities of M32
stars were elevated at large radii because of the presence of a dark
halo, then fitting a constant mass-to-light ratio model should yield
{\it higher} $M/L$ values when data at increasing radii are
included. We find instead the opposite. This strengthens the
conclusion that the integrated-light data provide no strong indication
for a dark halo in M32.

\subsection{Data-Model Comparison: Discrete Velocities}

\begin{figure*}\plotone{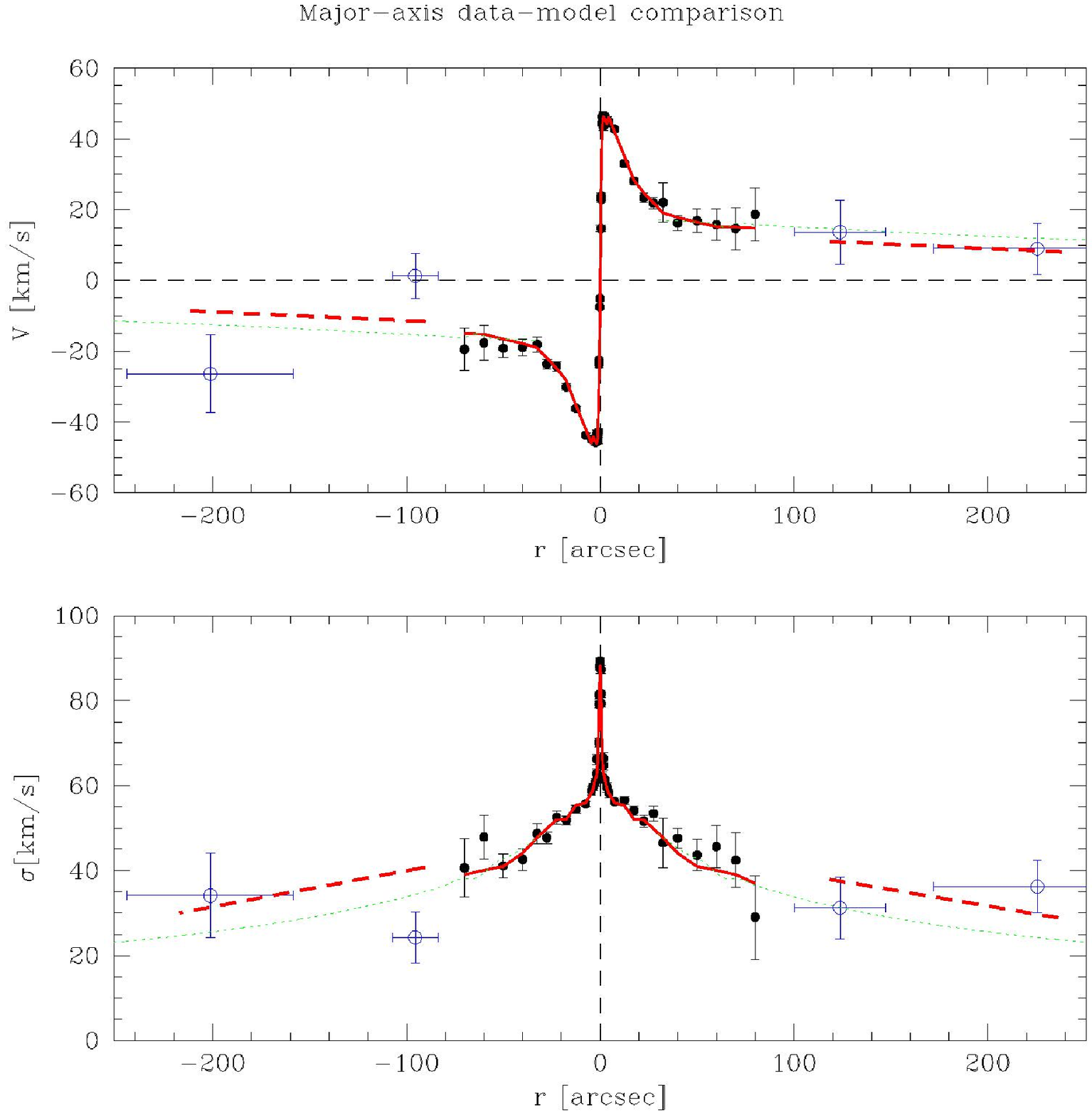}
\caption{Data-model comparison for the Gaussian parameters $V$ and
$\sigma$, along the major axis. Blue open data points are from fits to
the discrete velocity measurement histograms. Solid black data points
are from integrated-light measurements. Model curves shown in red display the predictions of the
best-fit three-integral axisymmetric edge-on model. Model predictions
for the integrated light data (corresponding to thin apertures) are
connected by a solid red line. Model predictions for the discrete
velocity data (corresponding to broad wedges, two on each side of the
galaxy) are connected by a dashed red line. For comparison, dotted
green curves at large radii show the major-axis predictions (not integrated over wedges) of a
two-integral model with the same geometry and mass-to-light ratio.}
\label{fig_plotmaj3}
\end{figure*}

\begin{figure*}\plotone{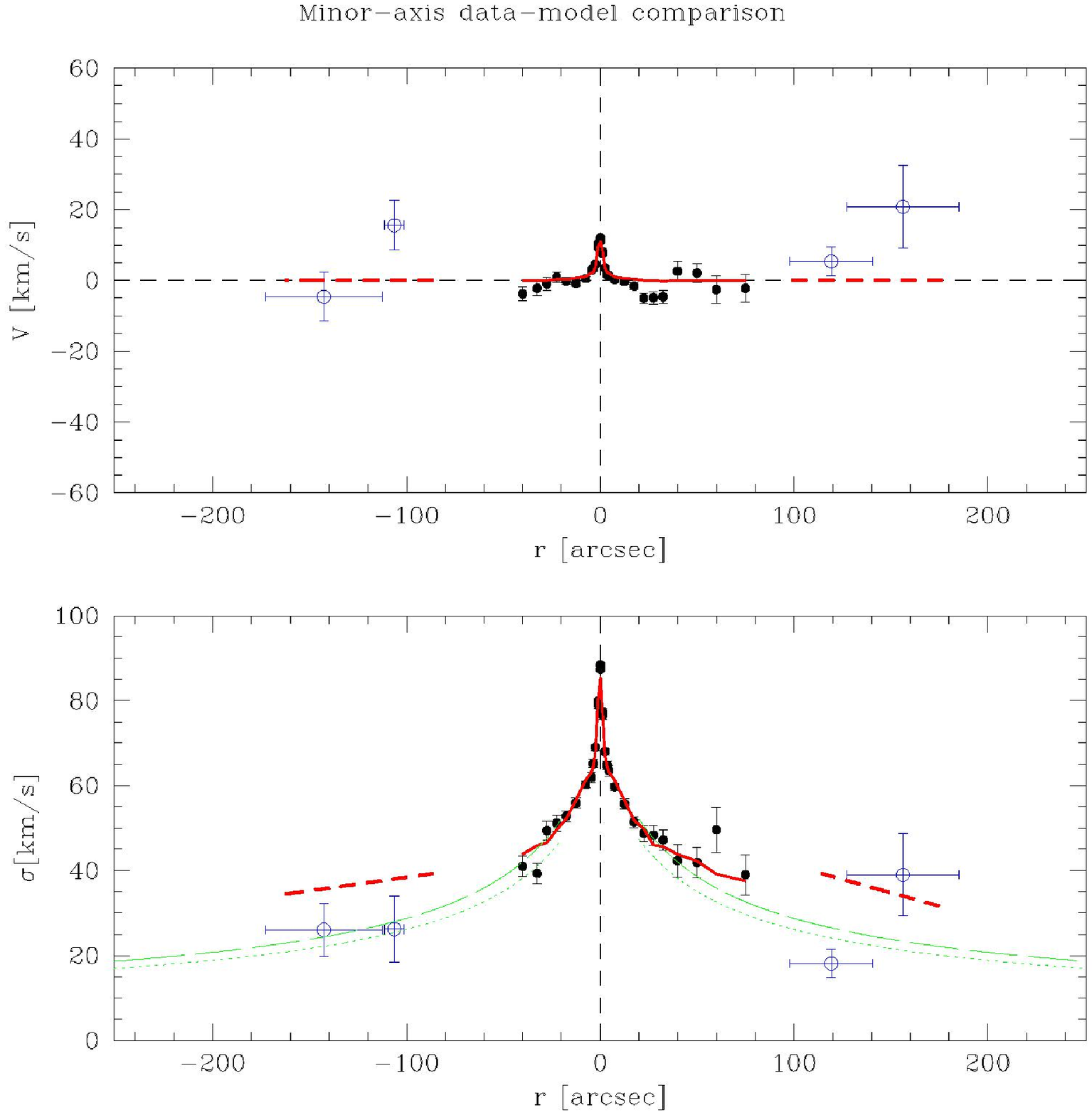}
\caption{Same as Figure~\ref{fig_plotmaj3}, but for the minor axis. The dotted green
curves at large radii show the predictions of a two-integral model
with the same geometry and mass-to-light ratio as the best-fit
three-integral model (red). The long-dashed green curves show the
predictions of a two-integral model with a 21\% larger mass-to-light
ratio.}
\label{fig_plotmin3}
\end{figure*}

The predictions of the best-fit model are compared to the $V$ and
$\sigma$ inferred from the discrete velocity measurements (blue open
dots) in Figures~\ref{fig_plotmaj3} and \ref{fig_plotmin3}, for the major and minor axis
respectively. The long-slit data (black solid dots) are also included
for comparison.

A three-integral orbit superposition model has the freedom to change
its dynamical structure with radius. It is therefore worthwhile to
consider for comparison simpler models that do not have this
freedom. The green curves in Figure~\ref{fig_plotmaj3} show the large
radii predictions of two-integral models with a distribution function
of the form $f(E,L_z)$. The Jeans equations of hydrostatic equilibrium
can be explicitly solved for such models, making them a simple
starting point for many analyses. They have been successfully applied
to the case of M32 in many prior studies, and here we have used the
same software as in \citet{van94a} to calculate their predictions. The
dotted green curves are for the same geometry and mass-to-light ratio
as in our best-fitting orbit superposition model.  We adopted a
parameter $k=0.57$ \citep[defined in e.g.][]{van94a} to split the
azimuthal motion into ordered and random components.

On the major axis, in the area around the ends of our long slits
(40\arcsec--100\arcsec\, from the center), the two-integral model predictions
are very similar to those from our best-fitting three-integral model
(solid red curves). On the minor axis, the three-integral model
predictions for the dispersion are higher than for the two-integral
model. This may be because: (a) the best-fitting three-integral model
has a dynamical structure that differs from a two-integral model; or
(b) the three-integral model predictions are obtained from Gaussian
fits to model LOSVDs, whereas the two-integral model predictions are
true dispersions. The long-dashed green curves in Figure~\ref{fig_plotmin3} show the
minor-axis predictions for a two-integral model with a 21\% higher
value of $M/L$. These appear more similar to the three-integral model
predictions.

The purpose of the green two-integral curves in
Figures~\ref{fig_plotmaj3} and \ref{fig_plotmin3} is to show how the
kinematics fall with radius in a constant mass-to-light ratio model in
which the dynamical structure itself does not vary with radius. The
gradient at large radii should not depend much on the dynamical
structure itself, as long as it is independent of radius. The discrete
velocity data points (blue points) at the large radii do {\it not}
follow this nominal behavior. In particular, the velocity dispersions
in the four outermost data points are higher than expected. The
two-integral models (using the higher $M/L$ on the minor axis) predict
on average $\sigma = 24.7$\kms\ at these radii. By contrast, the
observed weighted average dispersion is $33.0 \pm 3.5$\kms, which is
higher by a statistically significant $2.4\sigma$. This suggests an
increasing $M/L$ with radius (i.e., the presence of a dark halo),
unless the dynamical structure of M32 changes with radius beyond the
edges of our long slits.

The three-integral orbit superposition modeling approach automatically
adjusts the dynamical structure as necessary to best fit all available
data. The red dashed curves in Figures~\ref{fig_plotmaj3} and
\ref{fig_plotmin3} show the predictions thus obtained. For each side
of the galaxy (major or minor axis, positive or negative radius) there
are two data points. The predictions for these data points are
connected by a straight line. These predictions correspond to averages
over broad wedges on the sky, and not small apertures as was the case
for the integrated-light predictions. This affects primarily the
rotation velocity, which is smaller when averaged over a wedge than on
the major axis itself.

Given the significant scatter between the discrete velocity data
points, the three-integral model fits the data reasonably well. In
particular, the model predictions significantly exceed the nominal
model fall-off indicated by the two-integral models. For the four
outermost data points the predicted dispersion is 32.7\kms, consistent
with the weighted average $33.0 \pm 3.5$\kms\ of the observed values. For
the four inner data points the predicted dispersion does not fit the
average of the observed values, $38.6$\kms\ versus $24.9 \pm 2.3$\kms,
respectively. Apparently, a constant mass-to-light ratio model cannot
simultaneously reproduce the low dispersions observed in the inner
wedges of Figure~\ref{fig_stars}, while also reproducing the higher dispersions
observed in the outer wedges.

The reason that the three-integral models predict higher dispersions
than the two-integral models is due to a change in its dynamical
structure. Inspection of the dynamical structure of the best-fitting
three-integral model shows that it has increasing tangential
anisotropy towards larger radii. This causes more motion to be
observed along the line of sight direction. Moreover, models with
tangential anisotropy tend to have flat-topped LOSVDs. For such
LOSVDs, the dispersion of the best-fitting Gaussian (which is the
observed quantity) exceeds the true dispersion \citep{van93}. Both of
these effects help the model to fit the observed dispersions at large
radii.

\begin{figure}\includegraphics[trim=10 280 0 0, scale=0.85]{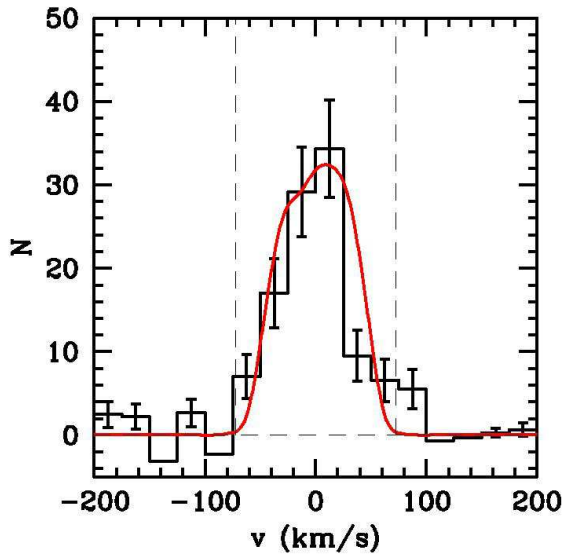}
\caption{Observed grand-total velocity histogram for the outer four
wedges in Figure~\ref{fig_stars}, with the M31 contribution (held fixed at the
values in Figure~\ref{fig_ml}) subtracted. The red curve is the prediction of the
best-fitting three-integral model. The horizontal axis shows the
velocity in the M32 frame ($v_{\rm hel} + 200$ \kms). Vertical dashed
curves at $\pm 72$ \kms indicate the escape velocity of the model at
$R=200\arcsec$ in the equatorial plane. The slight mismatch between the
observed and predicted histograms may be interpreted either as
evidence for an M32 dark halo, evidence for tidal perturbations in the
M32 outskirts, or evidence for velocity substructure in the M31 halo.}
\label{fig_histouter}
\end{figure}

To assess whether the tangential anisotropy of the best-fitting
three-integral model is consistent with the data it is necessary to
measure the shape of the LOSVD at large radii \citep{car95}. On
average, the outer wedges in Figure~\ref{fig_stars} each have only some 28 observed
M32 stars in them, much too little to reliable determine the
Gauss-Hermite moments of the LOSVD. Nonetheless, some important LOSVD
shape information can be obtained from the data. Figure~\ref{fig_histouter} shows the
observed grand-total velocity histogram for the outer four wedges,
with the M31 contribution (held fixed at the values in Figure 6)
subtracted. The red curve is the prediction of the best-fitting
three-integral model. The curves have similar widths, as was already
clear from the preceding discussion. However, there are some subtle
differences in the observed and predicted LOSVD {\it shape}.

The vertical dashed curves in Figure~\ref{fig_histouter} show the escape velocity of
the model at $R=200\arcsec$ in the equatorial plane (72\kms). The wings of
the predicted LOSVD fall to zero around this velocity. To achieve a
significant dispersion, the model creats a relatively flat-topped
LOSVD within the regime bounded by the escape velocity. By contrast,
the observed LOSVD histogram has a narrower core, and broader wings.
In particular, there are $\sim 6$ stars on the positive velocity side
of the LOSVD that move faster than the model escape velocity. If these
stars are bound to M32, then M32 must have a dark halo.  

However, alternative interpretations do exist. First, our simple model
for the LOSVD of the M31 halo is a smooth Gaussian. If in reality the
M31 halo has significant substructure (in velocity space) over the
region surrounding M32, then this result may cease to be
significant. It is then possible that the excess stars at $v \approx
95$\kms\ in the M32 frame (i.e., $v_{\rm hel} \approx -105$\kms) may
simply be a co-moving group of M31 stars. Second, it is possible that
we have reached a regime in M32 where tidal perturbations are playing
a role. In this case, it would not be appropriate to interpret the
excess stars in the context of an equilibrium model. The fact that the
observed histogram in Figure~\ref{fig_histouter} is not symmetric (there are excess
stars only on the positive velocity side) seems more consistent with
either of these interpretations than with evidence for a dark halo.

Other than the high-velocity tail of stars in Figure~\ref{fig_histouter}, our
data-model comparison provides very little evidence that the
kinematics of M32 might be affected by tidal perturbations or
non-equilibrium dynamics. The integrated-light measurements inside
$100\arcsec$ have a smooth behavior, and follow closely the
predictions of our equilibrium dynamical models. The discrete
measurements at larger radii show significant scatter, but this is
likely due to measurement errors, and not tidally induced. The
velocity dispersion increases from the inner to the outer wedges on
Figure~\ref{fig_stars}, but this increase is similar on all four sides of the
galaxy, and therefore not easily attributed to tidal perturbations.

In summary, by obtaining kinematics out to $\sim 8 r_{\rm eff}$ we
appear to have reached for the first time in M32 a regime where the
observed velocity dispersion flattens out as a function of radius. This can
be plausible interpreted as evidence for a dark halo (although we have
not actually constructed models with a dark halo explictly). On the
other hand, a three-integral constant-$M/L$ model with increasing
tangential anisotropy towards large radii can still fit all the data,
apart from half-a-dozen stars at $v_{\rm hel} \approx -105$
\kms. While these stars cannot be explained by any equilibrium model
without a dark halo, they may alternatively be due to substructure in
the M31 halo, or tidal perturbations in the M32 outskirts.

We have not explicitly explored dynamical models that include a dark
halo component. Given the added freedom of a mass-to-light ratio that
varies with radius, such models should certainly be able to fit the
data. However, given the limited number of data points at the largest
radii, it is unlikely that such models would be able to place strong
constraints on the properties of the dark halo.

   \section{SUMMARY \& DISCUSSION}\label{sec_discussion}

We have presented mean velocity, velocity dispersion and higher order Gauss-Hermite moment profiles along M32's major and minor axes based on Keck/DEIMOS spectroscopic observations of the integrated light and the resolved stellar population.  This study is the first to continuously transition between integrated light and the resolved stellar population in M32, or indeed in any galaxy.  The kinematical profiles provide the most radially extensive spectroscopic coverage for any cE galaxy, with measurements of the resolved stellar population extending to a projected distance of $\sim8~r^{\rm{eff}}_{I}$.  

We have constructed axisymmetric three-integral dynamical models for
M32 to interpret the new data. The integrated-light data out to $\sim
100\arcsec$ show falling dispersions with radius, which are well-fit by a
constant $M/L$ model. The discrete velocity data between 100--200\arcsec\ 
reveal a regime where the observed velocity dispersion flattens out as a function
of radius. This can be plausible interpreted as evidence
for a dark halo. However, a constant $M/L$ model can fit all the
available data out to the largest radius probed, provided that there
is increasing tangential anisotropy with radius in M32. The number of
observed M32 stars at large radii is too small to directly constrain
the anisotropy. A small number of fast moving stars at large radii
suggests that a dark halo may better explain the data than tangential
anisotropy, but these stars may also be due to substructure in the M31
halo, or tidal perturbations in the M32 outskirts.

It has long been known that M32's isophotes undergo a sharp twist and increase in ellipticity coincident with a break in the surface brightness profile at $\sim 5~r^{\rm{eff}}_{I}$, plausibly as a result of tidal interaction with M31 \citep[e.g.][]{ken87,cho02,joh02}.  Our kinematical study, however, shows no corresponding tidal signature across this region.  On the contrary, 
M32's kinematics appear to be symmetric and regular out to the limit of our survey.  
The lack of a strong gradient in the velocity and velocity dispersion profiles across the radius at which isophotal distortion occurs does not necessarily rule out the tidal distortion hypothesis.  
An alternative explanation is that the observed isophotal distortions in M32 are intrinsic to its structure.  \citet{fas89} find an increase in ellipticity and isophotal twisting in the outer regions of about half of all isolated elliptical galaxies, and conclude that tides are unlikely to be the cause of the distortion seen in these galaxies.  

M32 is a compact elliptical galaxy, and such galaxies are normally
found as satellites of more massive parent galaxies. As discussed in
\S\,\ref{intro_sec}, tidal stripping is a leading theory for their
formation. 
Our findings provide little support for a scenario in which tidal forces having significantly altered the M32 structure at large radii for two different reasons, although our conclusions are not strong enough to rule the tidal stripping model out all together.
First, the observed
kinematics provide little direct evidence of being impacted by tidal
effects. This appears surprising if the small size of M32 were
directly attributable to such effects. 
Second, the discrete velocity
measurements imply a relatively flat velocity dispersion profile at
large radii. This suggests
that M32
may well have a dark halo at radii of 200\arcsec\ and beyond, where
there is very little luminous matter. Dark and luminous matter are
both collision-less and subject to the same gravitational potential,
and tidal stripping removes material at large radii first. So if a
dark halo was left prominent at radii of 200\arcsec\ and beyond, then
it may be difficult to explain why the luminous body of the  galaxy got
stripped down to a radius $r_{\rm eff} \approx 30\arcsec$.

Neither the observed kinematics nor the dynamical modeling implications about the presence of a dark halo provide evidence that tidal forces have significantly altered the M32 structure at large radii.  These findings alone do not rule out tidal models for cE formation but instead suggest that tidal forces may not be as important in shaping cE galaxies as often argued.

   \acknowledgments 

KMH was supported in part by the Lawrence Scholars Program at Lawrence
Livermore National Laboratory (LLNL-JRNL-496754).  EK and KG were
supported through Hubble Fellowship grants 51256.01 and 51273.01,
respectively, from the Space Telescope Science Institute, which is operated
by the Association of Universities for Research in Astronomy, Inc., for NASA,
under contract NAS 5-26555.  PG, KMH, and BY acknowledge support from NSF
grants AST-0607852 and AST-1010039.  JSK's research is supported in part by a
grant from the STScI Director's Discretionary Research Fund.  KMH thanks
STScI and Yale University for their hospitality during her visits to carry
out some of this work.

The authors wish to recognize and acknowledge the very significant cultural
role and reverence that the summit of Mauna Kea has always had within the
indigenous Hawaiian community.  We are most fortunate to have the opportunity
to conduct observations from this mountain.

Facility: Keck:II (DEIMOS)

\clearpage


\begin{thebibliography}{57}
\bibitem[{{Bekki} {et~al.}(2001)}]{bek01}
{Bekki}, K., {Couch}, W.~J., {Drinkwater}, M.~J. \& {Gregg}, M.~D. 2001, \apj, 557, L42


\bibitem[{{Bender, Kormendy \& Dehnen}(1996)}]{ben96}
{Bender}, R., {Kormendy}, J. \& {Dehnen}, W. 1996, \apjl, 464, L123B


\bibitem[{{Bertin} {et~al.}(2002)}]{ber02}
{Bertin}, E., {Mellier}, Y., {Radovich}, M., {Missonnier}, G., {Didelon}, P., \& {Morin}, B. 2002, , in ASP Conf. Ser. 281, Astronomical Data Analysis Software and Systems XI, ed. D.~A.~Bohlender, D.~Durand, \& T.~H.~Handley (San Francisco: ASP), 228


\bibitem[{{Binggeli, Sandage \& Tammann}(1988)}]{bin88}
{Binggeli}, B. and {Sandage}, A. and {Tammann}, G.~A. 1988, \araa, 26, 509

\bibitem[{{Burkert}(1994)}]{bur94}
{Burkert}, A. 1994, \mnras, 266, 877

\bibitem[{Cappellari} {et~al.}(2006)]{cap06}
{Cappellari}, M., et al.  2006, \mnras, 366, 1126

\bibitem[{Cappellari} {et~al.}(2007)]{cap07}
{Cappellari}, M., et al.  2007, \mnras, 379, 418

\bibitem[{Carollo} {et~al.}(1995)]{car95}
{Carollo}, C.~M., {de Zeeuw}, P.~T., {van der Marel}, R.~P., {Danziger}, I.~J., \&
{Qian}, E.~E. 1995, \apj, 441, L25

\bibitem[{Carter \& Jenkins}(1993)]{car93}
{Carter}, D. and {Jenkins}, C.~R. 1993, \mnras, 263, 1049


\bibitem[{Chilingarian} {et~al.}(2009)]{chi09}
{Chilingarian}, I., {Cayatte}, V., {Revaz}, Y., {Dodonov}, S., {Durand}, D., {Durret}, F., {Micol}, A. \& {Slezak}, E. 2009, Science, 326, 1379

\bibitem[{Chilingarian \& Bergond}(2010)]{chi10}
{Chilingarian}, I.~V. \& {Bergond}, G. 2010, \mnras, 405, L11

\bibitem[{Choi} {et~al.}(2002)]{cho02}
{Choi}, P.~I., {Guhathakurta}, P. \& {Johnston}, K.~V. 2002, \aj, 124, 310

\bibitem[{Coelho, Mendes de Oliveira \& Cid Fernandes}(2009)]{coe09}
{Coelho}, P., {Mendes de Oliveira}, C. \& {Cid Fernandes}, R. 2009, \mnras, 396, 624

\bibitem[{Cretton} {et~al.}(1999)]{cre99}
{Cretton}, N., {de Zeeuw}, P.~T., {van der Marel}, R.~P., \& {Rix}, {H.-W.}
1999, \apjs, 124, 383


\bibitem[{Davidge}(2000)]{dav00a}
{Davidge}, T.~J. 2000, \pasp, 112, 1177 

\bibitem[{Davidge} {et~al.}(2000)]{dav00b}
{Davidge}, T.~J., {Rigaut}, F., {Chun}, M., {Brandner}, W., {Potter}, D., {Northcott}, M. \& {Graves}, J.~E. 2000, \apjl, 545, L89


\bibitem[{Dorman}{et~al.}(2012)]{dor12}
{Dorman}, C.~E., {Guhathakurta}, P., {Fardal}, M.~A., {Lang}, D., {Geha}, M.~C.,
{Howley}, K.~M., {Kalirai}, J.~S., {Cuillandre}, J.-C., {Dalcanton}, J.,
{Gilbert}, K.~M., {Seth}, A.~C., {Williams}, B.~F. \& {Yniguez}, B. 2012, \apj,
submitted

\bibitem[{Dressler \& Richstone}(1988)]{dre88}
{Dressler}, A. \& {Richstone}, D.~O. 1988, \apj, 324, 701D

\bibitem[{Drinkwater \& Gregg}(1998)]{dri98}
{Drinkwater}, M.~J. \& {Gregg}, M.~D. 1998, \mnras, 296, L15

\bibitem[{{Faber}(1973)}]{fab73}
{Faber}, S. M. 1973, \aj, 179, 423

\bibitem[{{Faber} {et~al.}(2003)}]{fab03}
{Faber}, S.~M., {et al.} 2003, Proc, SPIE, 4841, 1657

\bibitem[{{Falco} {et~al.}(1999)}]{fal99}
{Falco}, E.~E., {Kurtz}, M.~J., {Geller}, M.~J., {Huchra}, J.~P., {Peters}, J., {Berlind}, P., {Mink}, D.~J., {Tokarz}, S.~P. \& {Elwell}, B. 1999, \pasp, 111, 438

\bibitem[{{Fasano \& Bonoli}(1989)}]{fas89}
{Fasano}, G. \& {Bonoli}, C. 1989, \aaps, 79, 291


\bibitem[{Ford, Jacoby \& Jenner}(1978)]{for78}
{Ford}, H.~C., {Jacoby}, G.~H., \& Jenner, D.~C. 1978, \apj, 223, 94




\bibitem[{{Gilbert} {et~al.}(2007)}]{gil07}
{Gilbert}, K.~M., {Fardal}, M., {Kalirai}, J.~S., {Guhathakurta}, P., {Geha}, M.~C., {Isler}, J., {Majewski}, S.~R., {Ostheimer}, J.~C., {Patterson}, R.~J., {Reitzel}, D.~B., {Kirby}, E. \& {Cooper}, M.~C. 2007, \apj, 668, 245	
	


\bibitem[{{Goodman \& Lee}(1989)}]{goo89}		
{Goodman}, J. \& {Lee}, H.~M. 1989, \apj, 337, 84	
	
\bibitem[{{Graham}(2002)}]{gra02}	
{Graham}, A.~W. 2002, \apj, 568, L13	
	
\bibitem[{{Guhathakurta} {et~al.}(2005)}]{guh05}
{Guhathakurta}, P., {Ostheimer}, J.~C., {Gilbert}, K.~M., {Rich}, R.~M., {Majewski}, S.~R., {Kalirai}, J.~S. , {Reitzel}, D.~B. \& {Patterson}, R.~J. 2005, preprint (astro-ph/0502366) 

\bibitem[{{Guhathakurta} {et~al.}(2006)}]{guh06}
{Guhathakurta}, P., {Rich}, R.~M., {Reitzel}, D.~B., {Cooper}, M.~C., {Gilbert}, K.~M., {Majewski}, S.~R., {Ostheimer}, J.~C., {Geha}, M.~C., {Johnston}, K.~V. \& {Patterson}, R.~J., 2006, \apj, 131, 2497

\bibitem[{{Howley} {et~al.}(2008)}]{how08}
{Howley}, K.~M., {Geha}, M., {Guhathakurta}, P., {Montgomery}, R.~M., {Laughlin}, G. \& {Johnston}, K.~V. 2008, \apj, 683, 722

\bibitem[{{Huxor} {et~al.}(2010)}]{hux10}
{Huxor}, A., {Phillipps}, S., {Price}, J. \& {Harniman}, R., 2010, ArXiv Astrophysics e-prints, arXiv:astro-ph/1009.3185

\bibitem[{{Irwin} {et~al.}(2005)}]{irw05}
{Irwin}, M.~J., {Ferguson}, A.~M.~N., {Ibata}, R.~A., {Lewis}, G.~F. \& {Tanvir}, N.~R., 2005, \apjl, 621, L105

\bibitem[{{Jensen} {et~al.}(2003)}]{jen03}
{Jensen}, J.~B., {Tonry}, J.~L., {Barris}, B.~J., {Thompson}, R.~I., {Liu}, M.~C., {Rieke}, M.~J., {Ajhar}, E.~A. \& {Blakeslee}, J.~P. 2003, \apj, 583, 712

\bibitem[{{Johnston, Choi \& Guhathakurta}(2002)}]{joh02}
{Johnston}, K.~V., {Choi}, P.~I. \& {Guhathakurta}, P. 2002, \aj, 124, 127

\bibitem[{{Joseph} {et~al.}(2001)}]{jos01}
{Joseph}, C.~L., {Merritt}, D., {Olling}, R., {Valluri}, M., {Bender}, R., {Bower}, G., {Danks}, A., {Gull}, T., {Hutchings}, J., {Kaiser}, M.~E., {Maran}, S., {Weistrop}, D., {Woodgate}, B., {Malumuth}, E., {Nelson}, C., {Plait}, P. \& {Lindler}, D. 2001, \apj, 550, 668

\bibitem[{{Karachentsev} {et~al.}(2004)}]{kar04}
{Karachentsev}, I.~D., {Karachentseva}, V.~E., {Huchtmeier}, W.~K. \& {Makarov}, D.~I. 2004, \aj, 127, 2031

\bibitem[{{Kent}(1987)}]{ken87}
{Kent}, S~M. 1987, \apj, 266, 562

\bibitem[{{King \& Kiser}(1973)}]{kin73}
{King}, I.~R. \& {Kiser}, J. 1973, \apj, 181, 27

\bibitem[{{Kormendy}(1985)}]{kor85}
{Kormendy}, J. 1985, \apj, 295, 73

\bibitem[{Lauer} {et~al.}(1998)]{lau98}
{Lauer}, T.~R., {Faber}, S.~M., {Ajhar}, E.~A., {Grillmair}, C.~J., \& {Scowen}, 
P.~A. 1998, \aj, 116, 2263



\bibitem[{{McConnachie} {et~al.}(2005)}]{mcc05}
{McConnachie}, A.~W., {Irwin}, M.~J., {Ferguson}, A.~M.~N., {Ibata}, R.~A.,
  {Lewis}, G.~F. \& {Tanvir}, N. 2005, \mnras, 356, 979


\bibitem[{{Mieske} {et~al.}(2005)}]{mie05}
{Mieske}, S., {Infante}, L., {Hilker}, M., {Hertling}, G., {Blakeslee}, J.~P., {Ben{\'{\i}}tez}, N., {Ford}, H. \& {Zekser}, K. 2005, \aap, 430, L25

\bibitem[{{Monachesi} {et~al.}(2011)}]{mon11}
{Monachesi}, A., {Trager}, S.~C., {Lauer}, T.~R., {Freedman}, W., {Dressler}, A., {Grillmair}, C. \& {Mighell}, K.~J. 2011, \apj, 727, 55

\bibitem[{{Nieto}(1990)}]{nie90}
{Nieto}, {J.-L.} 1990, in Dynamics and Interactions of Galaxies, ed. R. Wielen (Berlin: Springer), 258

\bibitem[{{Nieto \& Prugniel}(1987)}]{nie87}
{Nieto}, {J.-L.} \& {Prugniel}, P. 1987, \aap, 186, 30


\bibitem[{{O'Connell}(1980)}]{oco80}
{O'Connell}, R.~W. 1980, \apj, 236, 430


\bibitem[{{Pritchet \& van den Bergh}(1994)}]{pri94}
{Pritchet}, C.~J. \& {van den Bergh}, S. 1994, \aj, 107, 1730

\bibitem[{{Rix \& White}(1992)}]{rix92}
{Rix}, H.-W. \& {White}, S.~D.~M. 1992, \mnras, 254, 389

\bibitem[{{Rood}(1965)}]{roo65}
{Rood}, H.~J. 1965, \aj, 70, 689


\bibitem[{{Rose} {et~al.}(2005)}]{ros05}
{Rose}, J.~A., {Arimoto}, N., {Caldwell}, N., {Schiavon}, R.~P., {Vazdekis}, A. \& {Yamada}, Y. 2005, \aj, 129, 712


\bibitem[{{Simon \& Geha}(2007)}]{sim07}
{Simon}, J.~D. \& {Geha}, M. 2007, \apj, 670, 313


\bibitem[{{Smith Castelli} {et~al.}(2009)}]{smi09}
{Smith Castelli}, A.~V., {Faifer}, F.~R., {Bassino}, L.~P., {Romero}, G.~A., {Cellone}, S.~A. \& {Richtler}, T. 2009, BAAA, 52, 229S

\bibitem[{{Sohn} {et~al.}(2007)}]{soh07}
{Sohn}, S.~T., {Majewski}, S.~R., {Mu{\~n}oz}, R.~R., {Kunkel}, W.~E., {Johnston}, K.~V., {Ostheimer}, J.~C., {Guhathakurta}, P., {Patterson}, R.~J., {Siegel}, M.~H. \& {Cooper}, M.~C. 2007, \apj, 663, 960

\bibitem[{{Stetson}(1994)}]{ste94}
{Stetson}, P.~B. 1994, \pasp, 106, 250

\bibitem[{{Tonry}(1984)}]{ton84}
{Tonry}, J.~L. 1984, \apjl, 283, L27

\bibitem[{{Tonry}(1987)}]{ton87}
{Tonry}, J.~L. 1987, \apj, 322, 632

\bibitem[{{Tonry} {et~al.}(2001)}]{ton01}
{Tonry}, J.~L., {et~al.}, 2001, \apj, 546, 681

\bibitem[{{Tremaine} {et~al.}(2002)}]{tre02}
{Tremaine}, S., {Gebhardt}, K., {Bender}, R., {Bower}, G., {Dressler}, A., {Faber}, S.~M., {Filippenko}, A.~V., {Green}, R., {Grillmair}, C., {Ho}, L.~C., {Kormendy}, J., {Lauer}, T.~R., {Magorrian}, J., {Pinkney}, J. \& {Richstone}, D. 2002, \apj, 574, 740

\bibitem[{{Walterbos \& Kennicutt}(1987)}]{wal87}
{Walterbos}, R.~A.~M. \& {Kennicutt}, R.~C., Jr. 1987, \aaps, 69, 311

\bibitem[{Valluri} {et~al.}(2004)]{val04}
{Valluri}, M., {Merritt}, D., \& {Emsellem}, E. 2004, \apj, 602, 66

\bibitem[{{van den Bosch \& de Zeeuw}(2010)}]{van10}
{van den Bosch}, R.~C.~E. \& {de Zeeuw}, P.~T. 2010, \mnras, 401, 1770

\bibitem[{{van der Marel} {et~al.}(1998a)}]{van98a}
{van der Marel}, R.~P., {de Zeeuw}, P.~T., \& {Rix}, {H.-W.} 1998a, \apj, 488, 119

\bibitem[{{van der Marel} {et~al.}(1998b)}]{van98b}
{van der Marel}, R.~P., {Cretton}, N., {de Zeeuw}, P.~T. \& {Rix}, H.~W. 1998b, \apj, 493, 613

\bibitem[{{van der Marel} {et~al.}(1997)}]{van97}
{van der Marel}, R.~P., {de Zeeuw}, P.~T., {Rix}, {H.-W.} \& {Quinlan}, G.~D. 1997, Nature, 385, 610

\bibitem[{{van der Marel} (1994)}]{van94}
{van der Marel}, R.~P. 1994, \mnras, 270, 271

\bibitem[{{van der Marel} {et~al.}(1994a)}]{van94a}
{van der Marel}, R.~P., {Evans}, N.~W., {Rix}, {H.-W.}, {White}, S.~D.~M. \& {de Zeeuw}, T. 1994a, \mnras, 271, 99

\bibitem[{{van der Marel} {et~al.}(1994b)}]{van94b}
{van der Marel}, R.~P., {Rix}, H.~W., {Carter}, D., {Franx}, M., {White}, S.~D.~M. \& {de Zeeuw}, T. 1994b, \mnras, 268, 521

\bibitem[{{van der Marel \& Franx}(1993)}]{van93}
{van der Marel}, R.~P. \& {Franx}, M. 1993, \apj, 407, 525


\bibitem[{{Verolme} {et~al.}(2002)}]{ver02}
{Verolme}, E.~K., {Cappellari}, M., {Copin}, Y., {van der Marel}, R.~P., {Bacon}, R., {Bureau}, M., {Davies}, R.~L., {Miller}, B.~M. \& {de Zeeuw}, P.~T. 2002, \mnras, 335, 517

\bibitem[{{Wirth \& Gallagher}(1984)}]{wir84}
{Wirth}, A. \& {Gallagher}, III, J.~S. 1984, \apj, 282, 85


\bibitem[{{Ziegler \& Bender}(1998)}]{zie98}
{Ziegler}, B.~L. \& {Bender}, R. 1998, \aap, 330, 819

\end{thebibliography}
\end{document}